\documentclass{aa}  

\usepackage{graphicx}
\usepackage{txfonts}
\usepackage{comment}
\usepackage{placeins}
\usepackage[unicode=true,colorlinks,linkcolor=magenta,citecolor=blue]{hyperref}
\begin{document}

   \title{TIC\,441725813: A new bright hybrid sdB pulsator\\ with differential core/envelope rotation}
   \titlerunning{TIC\,441725813: A new bright hybrid sdB pulsator with differential core/envelope rotation}

   \author{Wenchao Su
          \inst{1}
          \and
          Stéphane Charpinet\inst{1}
          \and
          Marilyn Latour\inst{2}
          \and
          Weikai Zong\inst{3,4}
          \and
          Elizabeth M Green\inst{5}
          \and
          Gang Li\inst{1}          
          }

   \institute{Institut de Recherche en Astrophysique et Planétologie, CNRS, Université de Toulouse, CNES, 14 Avenue Edouard Belin, 31400 Toulouse, France\\
              \email{wenchao.su@irap.omp.eu}
              \and
              Institut für Astrophysik und Geophysik, Georg-August-Universität Göttingen, Friedrich-Hund-Platz 1, 37077 Göttingen, Germany
              \and
              Institute for Frontiers in Astronomy and Astrophysics, Beijing Normal University, Beijing~102206, China
              \and
              School of Physics and Astronomy, Beijing Normal University, Beijing 100875, China
              \and
              Steward Observatory, University of Arizona, 933 N. Cherry Avenue, Tucson, AZ 85721, USA
             }

   \date{Received xxxx xx, xxxx; accepted xxxx xx, xxxxx}

 
  \abstract
   {The Transiting Exoplanet Survey Satellite (TESS) performs high-precision photometry over almost the whole sky primarily in search of exoplanet transits. It also provides exquisite data to study stellar variability, in particular for pulsating hot B subdwarf (sdB) stars.}
   {We present the detailed analysis of a new hybrid (p- and g-mode) sdB pulsator, TIC\,441725813 (TYC 4427-1021-1), discovered and monitored by TESS over 670 days.}
   {The TESS light curves available for this star were analysed using prewhitening techniques to extract mode frequencies accurately. The pulsation spectrum is then interpreted through methods that include asymptotic period spacing relationships and rotational multiplets identification. We also exploited a high signal-to-noise ratio (S/N), low-resolution spectrum of TIC\,441725813 using grids of non-local thermodynamic equilibrium (NLTE) model atmospheres to derive its atmospheric parameters.}
   {The light curve analysis reveals that frequencies are mostly found in the g-mode region, but several p-modes are also detected, indicating that TIC\,441725813 is a hybrid sdB pulsator.  We identify 25 frequencies that can be associated to $\ell=1$ g-modes, 15 frequencies corresponding to $\ell=2$ g-modes, and 6 frequencies characteristic of p-modes. Interestingly, several frequency multiplets interpreted as rotational splittings of deep-probing g-modes indicate a slow rotation period of at least $85.3 \pm 3.6$ day, while splittings of mostly envelope-probing p-modes suggest a significantly shorter rotation period of $17.9 \pm 0.7$ day, which implies the core (mainly the helium mantle with possibly the deeper partially-mixed helium-burning core that it surrounds) rotates at least ~4.7 times slower than the envelope. The radial velocity curves indicate that TIC\,441725813 is in a close binary system with a low-luminosity companion, possibly a white dwarf. While elusive in the available TESS photometry, a low-frequency signal that would correspond to a period of $\sim 6.7$\,h is found, albeit at low S/N.
   Furthermore, we estimate the inclination angle is $\sim 60$$^\circ$ by two independent means.
   }
   {
   TIC\,441725813 is a particularly interesting sdB star whose envelope rotates faster than the core. We hypothesise that this might be caused by the effects of tidal interaction with a companion, although in the present case, the presence of such a companion will have to be further investigated. This analysis paves the way toward a more detailed seismic probing of TIC\,441725813 using optimisation techniques, which will be presented in a second paper.}


   \maketitle
%

\section{Introduction}
Hot B subdwarf (sdB) stars are a class of compact stars in the late stages of evolution, associated with the Extreme Horizontal Branch (EHB) in the H-R diagram \citep{Heber1986}.
These stars consist of a burning helium core surrounded by a thin hydrogen-rich envelope.
Their mass is usually as low as $\sim 0.47$\,$\mathrm{M}_{\odot}$, and they have high effective temperatures ($\mathrm{T_{\rm eff}}$ $\sim$ 22,000 $-$ 40,000\,K) and surface gravities ($\log g \sim 5.6 - 6.2$; \citealt{{Saffer1994}}).
Since the envelope of sdB stars is very thin (less than $\sim 0.02$\,$\mathrm{M}_{\odot}$), after core helium exhaustion (about $\sim 150$\,Myr; \citealt{Dorman1993}) they do not undergo the asymptotic giant branch, but instead become hotter as He-shell burning objects (associated to the sdO class) and ultimately reach the white dwarf cooling sequence.

When a sdB star evolves to a specific temperature range and the equilibrium state of the star is perturbed, non-radial oscillations can develop, causing the surface of the star to expand and contract periodically and resulting in periodic variations of the star's luminosity. 
These sdB stars with luminosity variations are called pulsating sdB (sdBV) stars. 
The luminosity variations, which contain information about the stellar interior, provide opportunities to explore the structure and properties of sdB stars using asteroseismology \citep{Brassard2001,Charpinet2000,Charpinet2002a,Charpinet2002b,Charpinet19}.
Depending on the type of non-radial pulsations detected, sdBV stars are divided into two main groups. The first group shows rapid oscillations in the outer layers of the star with pressure as the restoring force. These are low radial order pressure(p)-mode (acoustic) pulsations whose periods typically range from 80 to 600 seconds. These modes are driven by a classical $\kappa$-mechanism involving partial ionisation of iron-group elements in the so-called $Z$-bump region and enhanced by the accumulation of such elements in that region by radiative levitation \citep{Charpinet1996,Charpinet1997}.
These rapidly oscillating sdB stars are now commonly referred to as the $\mathrm{sdBV_r}$ stars, V361\,Hya stars, or EC14026 stars \citep{Kilkenny1997,Kilkenny2010} and are found at the hotter end of the EHB ($\mathrm{T_{\rm eff}} \sim$ 28,000 -- 36,000\,K).
The second group of sdB pulsators, called the $\mathrm{sdBV_s}$ stars, V1093\,Her stars, or PG1716 stars \citep{Green2003}, are cooler ($\mathrm{T_{\rm eff}} \sim$ 22,000 -- 30,000\,K) 
and show slow oscillations from the interior of the star with gravity as the restoring force (g modes). These are mid radial-order g-mode pulsations whose periods are typically distributed between 1 and 4 hours. These g modes are driven by the same $\kappa$-mechanism involving iron-group elements \citep{Fontaine2003,Jeffery2006,Jeffery2007}.
Within these two main classes of sdBV stars, a fraction shows both p- and g-mode pulsations and are referred to as the hybrid pulsators, also known as the $\mathrm{sdBV_{rs}}$ or DW\,Lyn stars \citep{Schuh2006}. Their effective temperature is typically between the aforementioned two groups, in the $\sim$ 28,000 -- 30,000\,K range \citep{Heber2016}.

The observed pulsations in sdB stars are typically resonant standing or progressive waves whose distribution inside the star must satisfy boundary conditions, leading to a discrete set of eigenfrequencies characterised by three integers ($\rm n,\ell,\rm m$). The number n usually represents the number of nodes along the stellar radius, while in directions perpendicular to the radius the waves are described by spherical harmonic functions, where the degree $\ell$ represents the total number of nodal lines on the sphere and the azimuthal index m represents the number of nodal lines along the longitude circle, which takes values from $-\ell$ to $+\ell$ for a total of $2\ell+1$ values (see, e.g., \citealt{unno89}).

The nature of p-mode pulsations is similar to that of acoustic waves, and their eigenfrequencies increase with increasing radial order n.
On the opposite, g-mode pulsations behave like gravitational internal waves, and their eigenfrequencies decrease with increasing radial order n.
When n becomes large enough, in the so-called asymptotic regime, when propagating in a homogeneous medium, consecutive p-modes become equally spaced in frequency, while consecutive g-modes pulsations reach equal period spacing. This pattern can however be perturbed when rapidly changing physical conditions exist inside a star (a steep chemical gradient for instance).  
If stars were not rotating at all, the $2\ell+1$ waves associated with the same n and $\ell$ but with different m-values would have exactly the same frequency, which is called the m-number degeneracy. When rotation is present, however, this degeneracy is lifted. One frequency is split into $2\ell+1$ frequency components (called a multiplet) for a stationary observer, which is referred to as rotational splitting. Spacing between split frequencies within a multiplet is directly connected to the rotation rate of the regions where the modes propagate.

Based on rotational splittings of stellar oscillations, internal rotation profile at different stellar evolutionary stages have been revealed. Notably, the data collected by space-based missions, such as Kepler (\citealt{Borucki2010Sci}) and the Transiting Exoplanet Survey Satellite (TESS; \citealt{Ricker2014}), have increased the sample of stars with measured internal rotation rates to an unprecedented level.

On the main sequence, earlier than F-type stars exhibit gravity modes. These include $\gamma$\,Doradus stars, spanning A to F types \citep[e.g.][]{Kaye1999}, and slowly pulsating B-type (SPB) stars \citep{Waelkens1991}. Fast internal rotation in these stars modifies the g-mode period spacings, which is no longer equally spaced \citealt[e.g.][]{VanReeth2015}. This behaviour facilitates the measurement of internal rotation rates at the upper boundary of convective cores using the so-called traditional approximation of rotation \citep[e.g.][]{Townsend2003, Bouabid2013, VanReeth2016, LiGang2020_611_stars, Pedersen2022}. When these internal rotation measurements are combined with surface rotation rates determined through p-mode splittings or surface modulations, it is found that the differential rotation within the radiative layers is smaller than $\sim 10\%$ \citep[e.g.][]{Kurtz2014, LiGang2020_611_stars}. Furthermore, the coupling between g modes in the radiative layers and inertial modes within the convective cores suggests that differential rotation between these two regions should also remain below $\sim 10 \%$ \citep{Saio2021}.

After a phase of near-rigid rotation throughout the main sequence, stars in the post-main-sequence stage display more pronounced differential rotations. The development of differential rotation was observed in several subgiant stars \citep{Deheuvels2014, Deheuvels2020}. These stars exhibit an increase in the core-to-envelope rotation ratio from 1 during the main sequence to approximately 20 at the beginning of the red-giant-branch (RGB) phase. Core rotation rates for hundreds of RGB stars have been reported \citep{Gehan2018, Kuszlewicz2023ApJ}, which appear to be independent of the central magnetic fields \citep{Li2022Nature, Li2023}. In contrast, the envelope rotation rates for RGB stars have been measured in only a dozen stars \citep{DiMauro2016, Triana2017}, and large discrepancies remain between various measurement methods because the envelopes expand significantly in the RGB phase. Additionally, the differential rotation rates of several core-helium-burning stars have been measured, revealing mild differential rotations after the RGB phase \citep{Deheuvels2015}.

These past few years, reports that 5 pulsating sdB stars could exhibit significant (more than 1.5 times) radial differential rotation have been published \citep{Foster2015,Reed2019,Baran_2019,Reed2020,Silvotti2022}. In these stars, the outer layers seem to rotate faster than the core, leading to an intriguing situation that it is tempting to associate to angular momentum transfer brought about by internal gravity waves in synchronised binary systems \citep{Goldreich1989}, but further investigation is needed to confirm that hypothesis. Besides those specific cases, internal rotation has been measured in several other sdB stars using asteroseismology, showing that single sdB stars or relatively wide binaries are generally very slow rotators, with rotation periods similar to those measured for the cores of red-clump stars (see \citealt{Charpinet2018}). Moreover, a pulsating sdB component of a tight binary system was shown to be tidally synchronised down to at least half of its radius \citep{Charpinet2008,VanGrootel2013}.

Finally, we recall that the detection of rotational splittings in pulsating white dwarfs also permitted to probe their internal rotation (e.g., \citealt{Charpinet2009,Hermes2017}), indicating that most white dwarfs are solid-body rotators with rotation rates compatible with conservation of angular momentum since the RGB phase (see \citealt{Charpinet2018} for a discussion of this). 

Before the advent of high-precision fast-photometry from space, detailed asteroseismic studies of pulsating sdB stars were essentially limited to $\mathrm{sdBV}_{\rm r}$ pulsators, due mostly to the lower duty cycle and shorter duration of ground-based observation campaigns, making it difficult to detect and resolve long period g-mode pulsations in these stars.
Nowadays, with large amounts of data available from space telescopes, pulsating sdB stars of all flavours have become targets of choice for asteroseismology.
In this work, we focus on TIC\,441725813 (also known as TYC 4427-1021-1, 2MASS J17045838+7304433, or Gaia DR3 1655107708129775744), which is one of the brightest known sdB stars (with $\rm V=10.90$; \citealt{Hog2000}) only classified as such very recently.
This star is located in the TESS Continuous Viewing Zone (CVZ) and was observed, by the time of writing, for more than 670 days with this instrument. In the following, we present the first part of a detailed analysis of TIC\,441725813
based on these high-quality photometric data and additional spectroscopic data we could gather for that object. In section 2, we present the spectroscopic and photometric properties of TIC\,441725813. In Section 3, we provide a detailed analysis of the pulsation content of this star, leading to the identification of some of the observed modes as well as information about the rotation of TIC\,441725813. We also discuss constraints on the inclination angle of the star and the detection of a weak, low-frequency signal found in the TESS data that may hint toward the presence of a companion.
Finally, we summarise our results and prospects in Section 4.

\begin{table*}
\centering
\caption{Journal of spectroscopic observations of TIC\,441725813 from the 2.3m Bok Telescope}
\label{jour_spectro}
\renewcommand{\arraystretch}{1.05}
\begin{tabular}{ccccccc}
\hline
\hline
\noalign{\smallskip}
UT Date   & UT Midpoint   & Exposure time  &  Airmass  &  S/N  &  S/N Blue       & Relative RV \\
          &               & (s)            &           &       &  (< 5000 \r{A}) & (km$\cdot$s$^{-1}$) \\

        \noalign{\smallskip}
        \hline
        \noalign{\smallskip}
2005-09-11  & 02:48:21.0   &    62   &  1.367  &  183.4 &  214.7   &    -53.05 \\
2007-09-10  & 02:44:48.9   &    56   &  1.359  &  178.1 &  208.1   &    -23.62 \\
2008-06-16  & 07:10:04.1   &    56   &  1.328  &  168.2 &  197.3   &      4.30 \\
2008-07-28  & 03:34:17.3   &    78   &  1.331  &  192.8 &  225.0   &     44.52 \\
2008-07-29  & 03:30:45.6   &    42   &  1.331  &  150.0 &  175.7   &     27.86 \\
\noalign{\smallskip}
\hline
\noalign{\smallskip}
\multicolumn{2}{c}{Shifted, combined spectrum} & 294  & 1.367 & 391.5 & 458.1  &   Std. dev. = 39.33 \\
\noalign{\smallskip}
\hline

\end{tabular}
\end{table*}

\section{Spectroscopic and photometric properties}

\subsection{Atmospheric parameters from spectroscopy}
Spectroscopic measurements are important to determine the effective temperature and surface gravity of a star. Such measurements are particularly critical for detailed seismic studies of pulsating sdB stars, in particular, to deal with degeneracy that can be encountered between potential seismic solutions \citep{Charpinet2005}.

Five spectra of TIC\,441725813 were available from former campaigns at the Steward Observatory 2.3m Bok Telescope on Kitt Peak, Arizona. Acquired between September 2005 and July 2008 (see Table~\ref{jour_spectro}), these spectra were obtained using an old Bok CCD with resolution 8.7\,\r{A} and cover a wavelength range from 3600\,\r{A} to 6900\,\r{A}. In order to combine the spectra to produce a single, high signal-to-noise ratio (S/N) spectrum, we had to velocity-shift each of them by the amount given in the last column of Table~\ref{jour_spectro}. It is interesting to note that the relatively large standard deviation of these radial velocity shifts ($\sim 39.3$\,km$\cdot$s$^{-1}$) suggests that TIC\,441725813 is likely a short-period spectroscopic binary (see Section 3.1 for further discussion of this). 
Since this star is bright, each acquired spectrum achieved a rather high S/N resulting in a combined spectrum with an S/N of 391 (458 in the blue, below 5000\,\r{A}).

The profiles of detected Balmer lines ($\rm{H_\alpha}$ to $\rm{H_{11}}$) and helium lines in this high-S/N combined spectrum were subsequently fitted using a grid of model atmospheres computed with the "Blanchette" composition. The Blanchette models are non-local thermodynamic equilibrium (NLTE) sdB atmospheric models with fixed metal abundances constructed from the results of \cite{Blanchette2008} and based on the public codes TLUSTY and SYNSPEC (\citealt{Brassard2010}, see sdB STAR part). The metal abundance pattern in these models is meant to represent a "typical" sdB star in a context where variations in the metal content are known to exist due to microscopic diffusion (competing gravitational settling, radiative levitation, and stellar winds).

   \begin{figure}
   \centering
   \includegraphics[width=9cm]{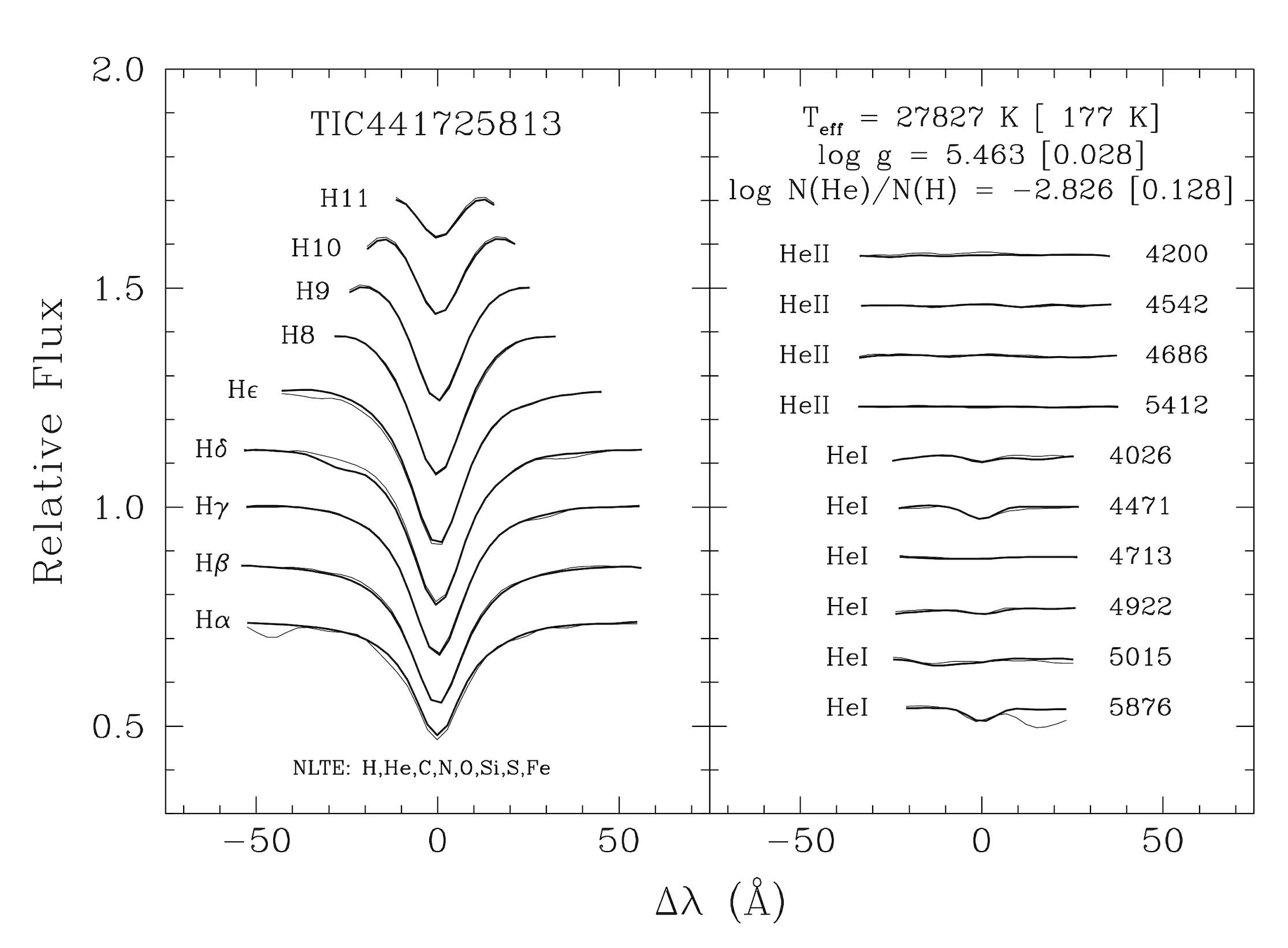}
      \caption{
      The optical spectrum of TIC\,441725813. All the available hydrogen and strong helium lines are shown as thin curves. Heavy curves are the model fit by using a 3D grid of NLTE synthetic spectra.}
         \label{spectrum}
   \end{figure}

The best fit obtained with these model atmospheres is shown in Fig.~\ref{spectrum} and leads to the following estimates for the atmospheric parameters of TIC\,441725813 : $\mathrm{T}_{\rm eff} = 27827\pm 177\,\mathrm{K}$, $\log g = 5.463\pm 0.028$, and $\log (\rm N_{\rm He}/\rm N_{\rm H}) = -2.826 \pm 0.128$. This star is therefore a rather cool sdB located in the $\log g - \mathrm{T_{\rm eff}}$ region where g-mode pulsators are usually found. 
These values are consistent (albeit slightly larger for $\mathrm{T_{\rm eff}}$ and $\log g$) with those reported in \cite{Baran_2024}, independently derived from the analysis of a spectrum obtained at the 2.56\,m Nordic Optical Telescope (NOT).

\subsection{TESS time series}
TESS -- the Transiting Exoplanet Survey Satellite -- is a satellite launched by NASA in 2018 to survey the whole sky. Its main scientific goal is to detect exoplanets transiting nearby stars \citep{Ricker2014}.
For this purpose, TESS is equipped with four identical custom wide-angle cameras, that together can monitor a 24-degree by 90-degree strip of the sky.
Each strip is nearly continuously observed for about 27 days and is called a "sector".
Due to overlap between sectors at high ecliptic latitudes, some stars can be observed longer. 
In particular, stars located in the continuous viewing zone (CVZ) can be monitored almost without interruption for more than 300 days during one cycle (roughly a year of TESS observation).
The TESS CCDs read out images continuously at 2-second intervals.
The data handling unit (DHU, a Space Micro Image Processing Computer in TESS) stacks the 2-second images in groups to produce the 20-second, 2-minute, or long-cadence\footnote{Throughout the core program and two extensions of the TESS mission, the cadence for the Full Frame Images has changed from 30 minutes originally to 10 minutes, and now 200 seconds.} (Full-Frame Images) observations. 
Besides the search for transits, the long duration, mostly uninterrupted, high-precision photometry of TESS allows to accurately extract periodic signals affecting stellar luminosity, including stellar pulsations.

\begin{figure}
   \centering
   \includegraphics[width=9cm]{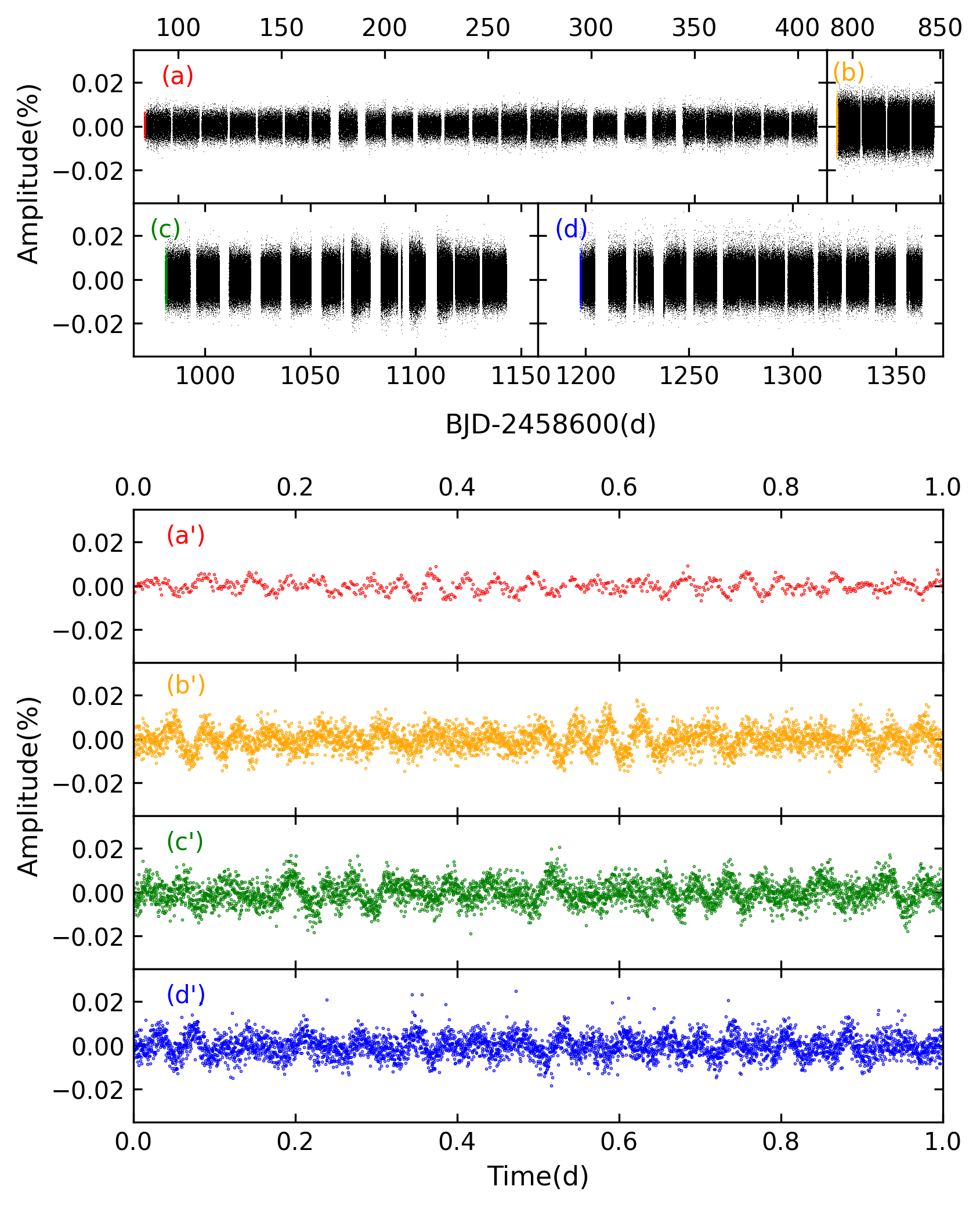}
      \caption{Photometry of the four data sets obtained by TESS for TIC\,441725813. The first set covers 12 sectors from S14 to S25 (only 2-min cadence data available; top panel (a)). The second, third, and fourth data sets consist of 20-s cadence data for S40-S41, S47-S52, and S55-S60 respectively (top panels (b), (c), and (d)). Taking the first data set as an example: the upper panel (a) shows the entire light curve spanning 312.1 days and sampled every 2 minutes. Gaps are due to the interruption during data download and points removed from the light curve because of a non-optimal quality flag warning. The bottom panel (a$'$) is a close-up view of the first day of observation, which shows clearly brightness variations caused by pulsations.
              }
         \label{lc_all}
   \end{figure}

As of this work, TIC\,441725813 has been observed in 26 sectors, with a cumulative observation time baseline of over 674 days.
Taking into account time gaps and available cadences of these observations, we divided the data into four groups for independent analyses.
The first data set includes 12 consecutive sectors (S14 to S25), in which TESS has continuously monitored TIC\,441725813 with the 2-min cadence mode for more than 300 days.
The second data set consists of S40 and S41 which cover about 55 days. 
Short-cadence observations (20s mode) became available at that time and could reveal higher-frequency signals. 
Therefore, we chose to use the 20s data when both cadences were available.
The remaining 12 sectors (S47 to S52, S55 to S60) make up the third and fourth data sets, each with a total time baseline of more than 150 days. For these two sets, we also chose to focus on the 20-s cadence data.
Details (start time, number of valid data points $\rm N$, and run duration $\rm{\delta T}$) concerning the four sets of TESS photometric data are provided in Table~\ref{jour_obs}.
Fig.~\ref{lc_all} shows the entire light curve and a close-up view on the first day for each set defined above. It can be seen that the first set of data has a very long time coverage with a good continuity, which is conducive to the discovery of more refined frequency structures.
The other sets of data are more densely sampled (with the 20s-cadence mode), offering the possibility to search for higher frequencies.
\begin{table}
\centering
\caption{Journal of TESS observations of TIC\,441725813}
\label{jour_obs}
\renewcommand{\arraystretch}{1.05}
\begin{tabular}{ccll}
\hline
\hline
\noalign{\smallskip}
Sect.   & BJD Start       & N                          & $\delta$T \\
        & (BJD-2457000)   &                            & (d)       \\ 
        \noalign{\smallskip}
        \hline
        \noalign{\smallskip}
14      & 1683.347        & 19337                      & 26.9      \\
15      & 1711.358        & 18757                      & 26.1      \\
16      & 1738.646        & 17765                      & 24.7      \\
17      & 1764.678        & 18012                      & 25.0      \\
18      & 1790.651        & 17554                      & 24.4      \\
19      & 1816.077        & 18052                      & 25.1      \\
20      & 1842.499        & 18954                      & 26.3      \\
21      & 1870.429        & 19694                      & 27.4      \\
22      & 1899.301        & 19579                      & 27.2      \\
23      & 1928.100        & 19279                      & 26.8      \\
24      & 1955.789        & 19074                      & 26.5      \\
25      & 1983.627        & 18489                      & 25.7      \\ 
\noalign{\smallskip}
\hline
\noalign{\smallskip}
40      & 2390.652        & 426409                     & 28.2      \\
41      & 2419.988        & 402059                     & 26.6      \\
\noalign{\smallskip}
\hline
\noalign{\smallskip}
47      & 2581.031        & 391776                     & 25.9      \\
48      & 2611.375        & 372141                     & 24.6      \\
49      & 2640.414        & 361417                     & 23.9      \\
50      & 2665.271        & 396746                     & 26.2      \\
51      & 2692.947        & 371767                     & 24.6      \\
52      & 2718.634        & 369562                     & 24.4      \\ 
\noalign{\smallskip}
\hline
\noalign{\smallskip}
55      & 2797.093        & 410726                     & 27.2      \\
56      & 2825.249        & 421583                     & 27.9      \\
57      & 2853.347        & 434865                     & 28.8      \\
58      & 2882.323        & 402252                     & 26.6      \\
59      & 2910.258        & 369527                     & 24.4      \\
60      & 2936.897        & 343501                     & 22.7      \\ 
\noalign{\smallskip}
\hline
\noalign{\smallskip}
\multicolumn{2}{c}{Total} & \multicolumn{1}{l}{5337460} & 674.1     \\ 
\noalign{\smallskip}
\hline
\end{tabular}
\end{table}

\section{Light curve analysis}
\begin{figure}
   \centering
   \includegraphics[width=9cm]{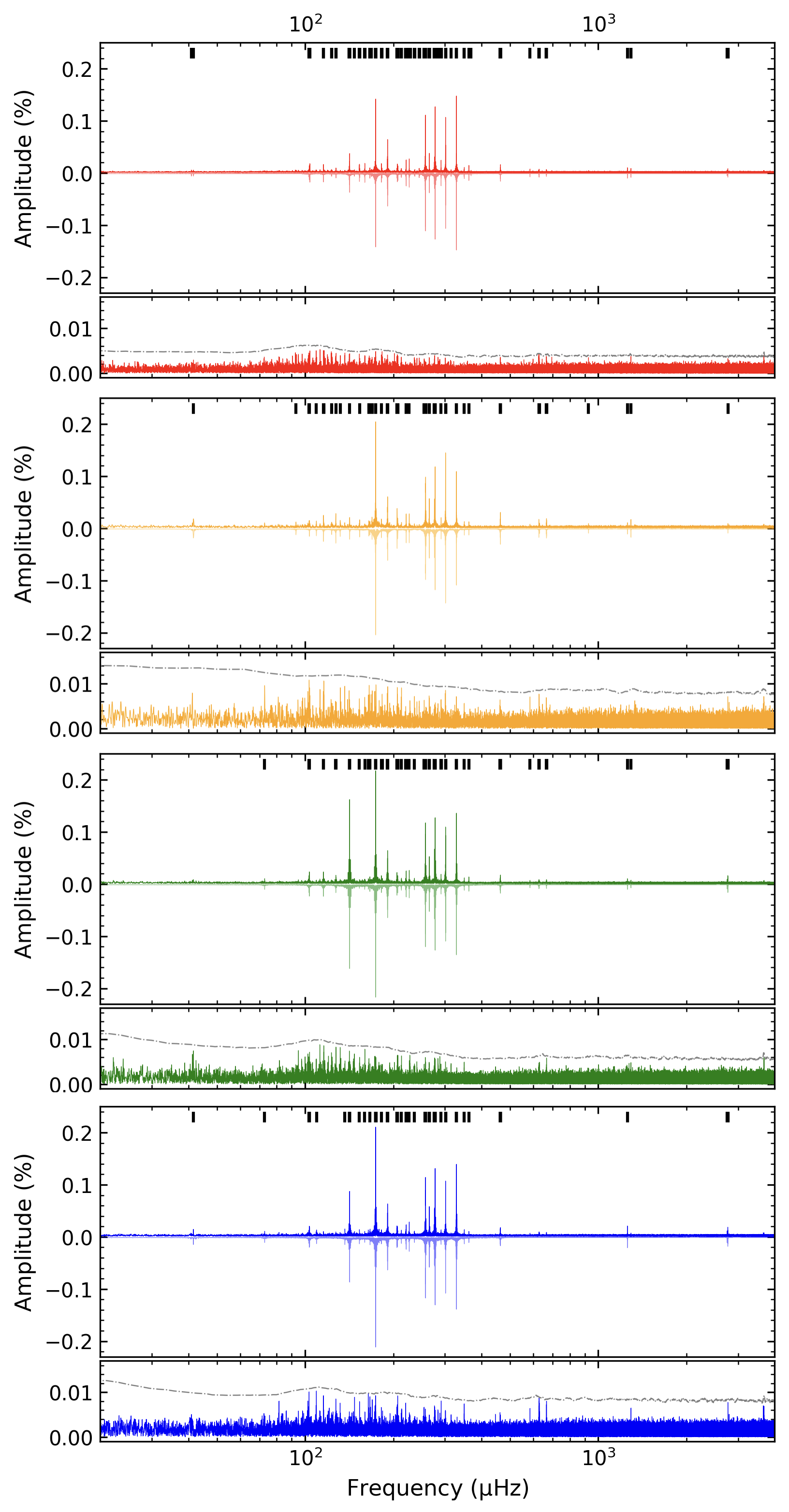}
      \caption{Lomb-Scargle Periodograms (LSP; amplitude in \% of the mean brightness vs. frequency in $\mu$Hz) for each data set constructed from the TESS observation of TIC\,441725813 (from top to bottom, respectively). The grey horizontal line indicates our conservative threshold of 5.2 times the median noise level (see text). The upper panel in each diagram displays all signals extracted above 5.2 times S/N. The curve plotted upside-down is the reconstruction of the LSP based on all extracted frequencies. The lower panels show the residual left after extracting all the significant frequencies.
              }
         \label{fre_amp}
   \end{figure}

We computed Lomb-Scargle Periodograms (LSP) and applied standard prewhitening and nonlinear least-square fitting techniques \citep{Deeming1975} implemented in the dedicated software FELIX (Frequency Extraction for LIghtcurve eXploitation; \citealt{Charpinet2010,Zong2016}) to extract the frequencies present in the light curve of each data set considered. 
Fig.~\ref{fre_amp} shows the most relevant sections (those with significant signals present) of the frequency spectra calculated for the four data sets. We placed our detection threshold at a conservative S/N value of 5.2 times the local median noise, following tests carried out in \citet{Charpinet19} and based on the method of \citet{Zong2016}. With this threshold, the False Alarm Probability (FAP) that a peak could be due to a noise fluctuation is less than 0.01\%. 
The data reveal that TIC\,441725813 is a rich sdB pulsator. In the first data set, we extracted a total of 172 frequencies, most of them concentrated in the g-mode pulsation region between 100 and 400\,$\mu$Hz. The residual indicates that even more signal must be present in that frequency range, but at amplitudes too low to confidently extract additional individual modes. We do not consider these further in the following. In addition to g-modes, significant peaks are also present in the p-mode region above 1000\,$\mu$Hz and, interestingly, two weak peaks are found in the region below 50 $\mu$Hz, a frequency range well below the g-mode frequency cut-off in typical sdB pulsators (see the supplementary material of \citealt{Charpinet2011a}) and where the signal is generally associated to orbital modulations. We discuss further the nature of those peaks in the following subsection.
We note that over the long time baseline of this data set, most pulsation modes are affected by long-term variations of their amplitude and/or phase,
often leading to residuals in the frequency prewhitening process that assumes stable periodic variations (see \citealt{Zong2016b}). Such residual frequencies, which were indiscriminately counted amongst the 172 frequencies mentioned earlier, should not be considered as independent modes. After grouping and discarding these residuals around main frequencies, keeping only the highest amplitude component as the real signal, we end up with 35 independent\footnote{In this context, independent refers to frequencies identified as real signals from pulsations or other astrophysical sources, excluding close residual frequencies that may result from signal modulations.}
frequencies listed in Appendix~\ref{fre_mi}\footnote{For completeness, all frequencies obtained from the raw extraction are provided in electronic form at the CDS via anonymous ftp to cdsarc.u-strasbg.fr (130.79.128.5) or via \href{http://cdsweb.u-strasbg.fr/cgi-bin/qcat?J/A+A/}{http://cdsweb.u-strasbg.fr/cgi-bin/qcat?J/A+A/}.}. 
The frequencies detected in the second data set are very similar to their counterparts from the first set, but due to the shorter observing time, the noise level is higher, and fewer frequencies are above the threshold (75 in total, but 27 independent frequencies given in Appendix~\ref{fre_mi}). For similar reasons, we extracted a total of 85 peaks in the third data set, resulting in a list of 30 independent frequencies. In the fourth data set, we identified 71 peaks, yielding 25 independent frequencies, as also detailed in Appendix~\ref{fre_mi}. Considering all results from the four data sets, we can first conclude that TIC\,441725813 clearly has both g- and p-mode pulsations and is, therefore, a newly identified hybrid sdBV star. In addition, one of the weak low-frequency signals at 41.6\,$\mu$Hz referred to previously is clearly present in at least three data sets, making it a robust detection. It shows however a puzzling modulation of its amplitude, with it being at its strongest in the second data set. The other weak peak at 40.9\,$\mu$Hz is only seen in the first data set with no counterpart below the detection threshold in the second and third sets and appears below the threshold in the fourth set, thus casting doubts about its reality.
We also found mode splittings that we associate with the star rotation for both g- and p-modes. We analyse them in more detail in the following subsections.

\subsection{Low-frequency signals}

\begin{figure}
   \centering
   \includegraphics[width=9cm]{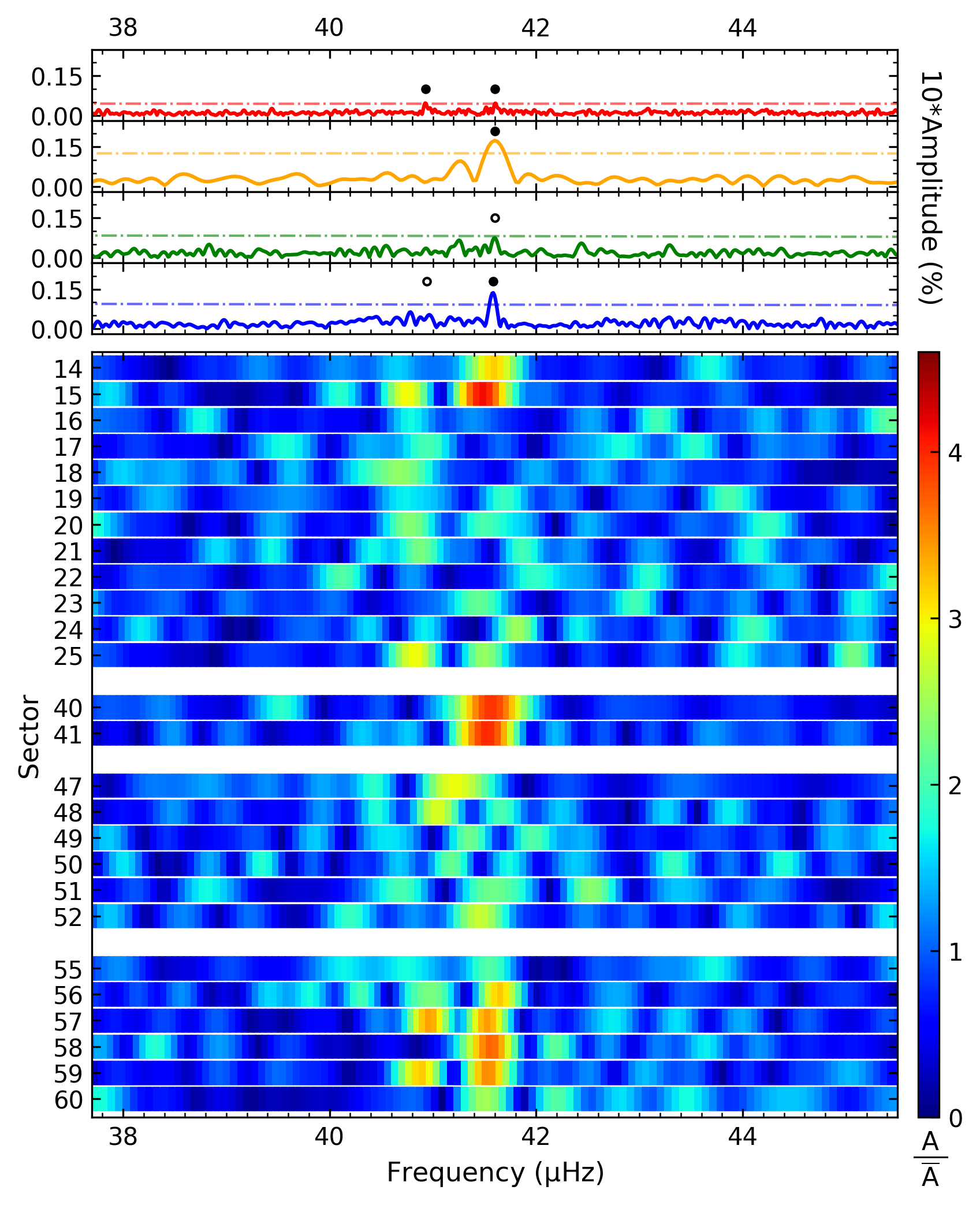}
      \caption{
      Close-up view of the Lomb-Scargle Periodograms in the low-frequency range.
      The upper panels show the LSP of the four data sets in that region. The red horizontal line indicates our adopted detection threshold of 5.2 times the local median noise level. The black dots mark the position of signals with S/N larger than 5.2. Black circles are possible signals just below the threshold. The lower panels show time-frequency maps built from the LSP of each sector in the same frequency range. Amplitudes are colour-encoded according to the
      scale shown in the colour bar on the right-hand side.
              }
         \label{stft_felix}
   \end{figure}

A close-up view of the low-frequency range in the Lomb-Scargle Periodograms of the four data sets is shown in the top panels of Fig.~\ref{stft_felix} (see also Appendix~\ref{fre_mi}). 
In the first data set, we extracted two close peaks with similar frequencies and amplitudes, at 40.92\,$\mu$Hz and 41.60\,$\mu$Hz, respectively. In the second set, we only found one peak with a frequency of 41.59\,$\mu$Hz, but no
signal at all near 40.92\,$\mu$Hz even below our detection threshold (the LSP is flat at this frequency position).
In the third set, we did not find any significant signal, strictly speaking, but a very weak peak just below the detection limit is noticeable at the 41.6\,$\mu$Hz location, while the signal around 40.92\,$\mu$Hz is still totally absent. In the fourth set, we once again extracted the significant signal at the 41.6\,$\mu$Hz location and there is the possibility of a very faint signal (S/N $\sim$ 3.3) below the detection limit at 40.92\,$\mu$Hz.
 
\begin{figure}
   \centering
   \includegraphics[width=9cm]{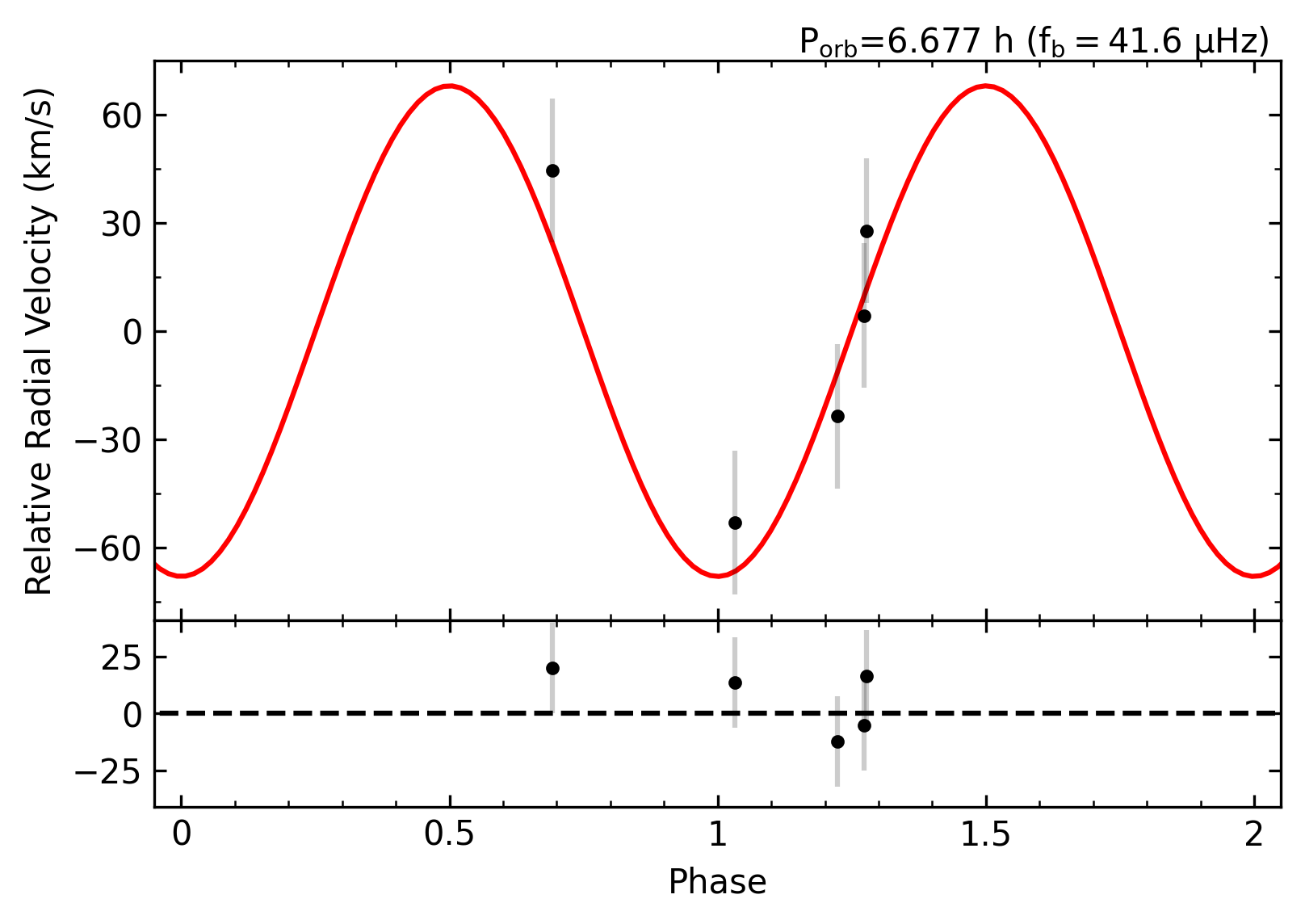}
      \caption{
      Phase - Radial velocity diagram. {\sl Upper panel}:  The red line represents the radial velocity curve assumed for the orbital signal at 41.6\,$\mu$Hz (6.677 h). The black dots represent the relative radial velocities obtained from spectroscopic observations, with gray vertical lines indicating estimated errors. {\sl Lower panel}: The residual plot of the fitting results is shown.
              }
         \label{phase_rv}
   \end{figure}

From these observations, we estimate that only one signal (the 41.60\,$\mu$Hz frequency) can be safely considered as real from its recurrent appearance (or near appearance) throughout the entire set of data.
In this low-frequency range, it is tantalizing to associate such a signal to an orbital modulation, as it corresponds to typical orbital timescales encountered in close binaries with an sdB component (in that case an orbital period of about 6.7h). Such an interpretation is reinforced by available spectroscopy suggesting that TIC\,441725813 may be a binary system due to the relatively large scatter in radial velocity for each individual spectrum (Section 2). We tested further this possibility with the results displayed in Fig.~\ref{phase_rv}. To build that radial velocity phase curve, we assumed that the period (6.667 h) corresponding to 41.6\,$\mu$Hz is indeed the orbital period of a binary system and we phased the relative radial velocity measurements provided in Table~\ref{jour_spectro} relative to that period. By chance (these spectra were taken at random times), the only five available measurements turn out to be favourably distributed in phase, leading to a highly suggestive match that reinforces the idea that TIC\,441725813 is indeed a binary system with that orbital period. Additional spectroscopy to fully sample the radial-velocity phase curve is however needed for a definitive conclusion on this. While the orbital interpretation for this signal is gaining impetus, the colour map in the bottom panel of Fig.~\ref{stft_felix} showing the LSP sector by sector indicates however that its amplitude seems to vary with time, where one would expect an orbital modulation to generate a variation of constant amplitude. Since this signal is very weak, we examined, with the pixel data, potential contamination from nearby stars that could lead to a spurious or inconsistent signal in the TESS photometry of TIC\,441725813. We found only two faint non-variable stars in close proximity to our target that cannot have any significant contribution. Therefore, the signal is confirmed to originate from TIC\,441725813 itself and part of its amplitude variations may be of intrinsic nature.
At this stage, however, we refrain from speculating further on this particular issue until the orbital origin of the low-frequency signal is confirmed with more radial velocity measurements. 

\subsection{Rotational Splittings}
When a pulsating star has a slow rotation, pulsation modes with the same values of $\ell$ and n, but different m split into $2\ell+1$ components which, to first-order approximation, are equally spaced in frequency. This spacing depends on a weighted average of the rotation rate of internal regions where the modes propagate. If higher-order terms are negligible (which is generally the case for the typically long rotation timescales encountered in sdB stars), the relation between the frequency splitting and the rotation rate can be expressed as \citep{Charpinet2018}:
\begin{equation}
\centering
    \Delta \nu_{n\ell} = \frac{1}{2\pi}\int_0^R K_{n\ell}(r)\Omega_{\mathrm{rot}}(r)\mathrm{d}r
    \label{eqn_splitting}
\end{equation}
where $\Delta \nu_{n\ell}$ is the spacing in frequency between modes of consecutive m-values ($|\Delta m|=1$), but same $\ell$ and n. $\Omega_{\rm rot}$ ($\equiv 2\pi/P_{rot}$) is the spherically symmetric rotation law (expressed in units of angular frequency), and $K_{n\ell}$ is the first-order rotation kernel for the mode considered, which acts as a weight function. Its general form is given by
\begin{equation}
\centering
\label{kernel}
K_{n\ell}(r)= \frac{\xi^2_{r}+\left[\ell(\ell+1)-1\right]\xi^2_{h}-2\xi_{r}\xi_{h}}{\int_0^R\left[\xi^2_{r}+\ell(\ell+1)\xi^2_{h}\right]\rho r^2 \mathrm{d}r}\rho r^2
\end{equation}
where $\xi_{r}$ and $\xi_{h}$ are, respectively, the real parts of the radial and horizontal components of the displacement vector (see \citealt{unno89}).
\begin{equation}
\centering
\label{dv}
\Vec{\xi}=\left[\xi_{r}(r)\;,\;\xi_{h}(r)\frac{\partial}{\partial \theta}\;,\;\xi_{h}(r)\frac{\partial}{\partial \theta}\frac{1}{\sin\theta}\right]Y^{m}_{\ell}(\theta,\phi)e^{i\sigma t}.
\end{equation}

Eq.~\ref{eqn_splitting} implies that the frequency splitting reflects mainly the rotation rate in stellar layers where the kernel is large, as parts of the star where this kernel is close to zero do not contribute much to the integral value. This means that different types of modes, with significantly different kernels, can provide localised information on the rotation rate inside the star. This property has indeed been used in asteroseismology to invert internal rotation profiles, mainly in the Sun or Sun-like stars (see \citealt{schunker16} and references therein), with techniques such as the Optimally Localised Averaging Kernels (OLA) method \citep{backus68}. Without going into that much sophistication (these methods require the detection of many rotational multiplets to be applied), the presence in hybrid sdB pulsators of both p- and g-modes carries an interesting potential to easily extract localised information on the rotation rate in these stars. 

\begin{figure}
   \centering
   \includegraphics[width=9cm]{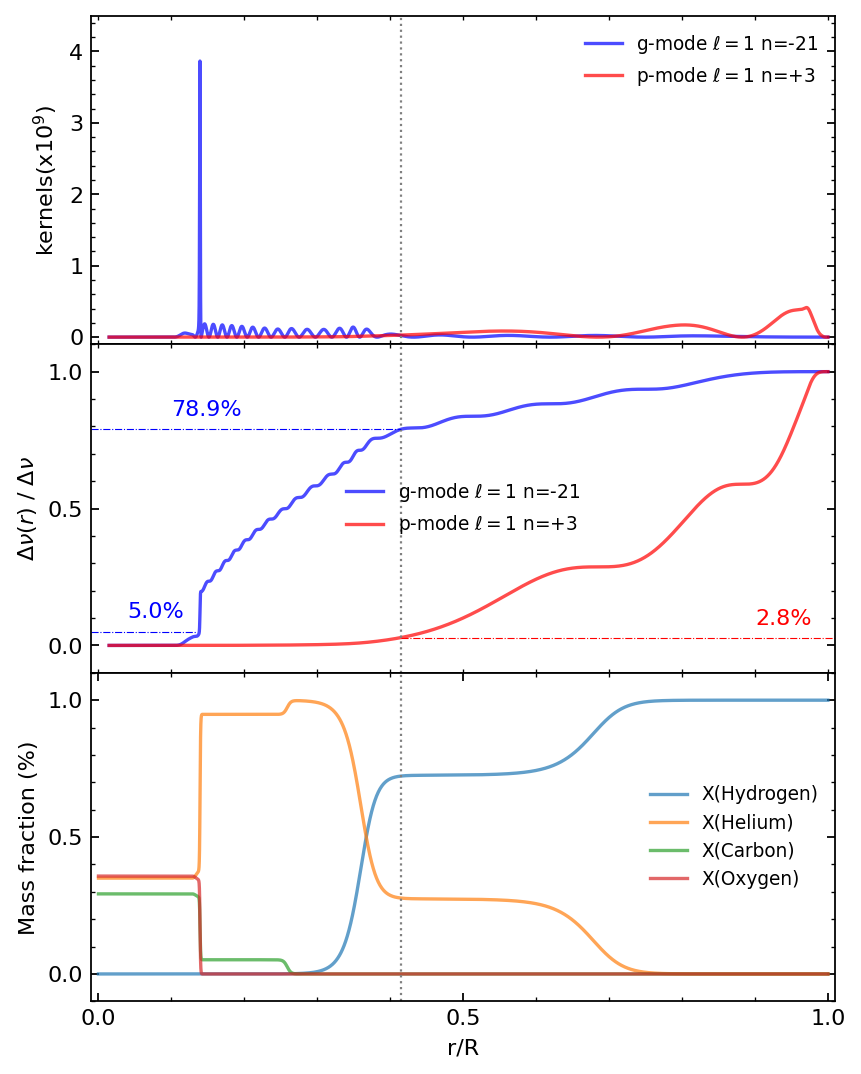}
      \caption{{\sl Top panel}: Rotation kernel as a function of normalised radius for a typical g-mode (blue line; $\ell=1$, $\rm n=-21$) and a typical p-mode (red line; $\ell=1$, $\rm n=3$) in our representative model of TIC\,441725813. {\sl Middle panel}: Running integral of Eq.~\ref{eqn_splitting} normalised to one showing which stellar regions contribute most to the frequency splitting for each mode. {\sl Bottom panel}: Chemical stratification in our adopted representative model showing the run of the hydrogen, helium, carbon, and oxygen mass fractions. The vertical line present in all panels indicates the boundary between the hydrogen-rich envelope and the helium mantle.
              }
         \label{kernels}
   \end{figure}

This is illustrated in Fig.~\ref{kernels} that compares two representative kernels from, respectively, a typical g-mode and a typical p-mode. To compute these kernels, we calculated, with our codes STELUM and PULSE (see \citealt{Charpinet19} and references therein), a static sdB stellar model and its pulsation properties representative of TIC\,441725813 in that its $\mathrm{T}_{\rm eff}$ and $\log g$ (27,780~K and 5.339~dex, respectively) are close to the values derived for that star from spectroscopy (see Section 2). We emphasise that this model does not constitute a seismic model of TIC\,441725813 per se, as its other parameters (e.g., the mass set to 0.47 $\mathrm{M_{\odot}}$) and internal structure were fixed to canonical or typical values with no attempt to match the observed pulsation periods (the latter is left to the second paper of this series). The objective here is to provide insight into the general properties of modes in sdB stars that do not depend much on the details of the model. In Fig.~\ref{kernels}, the bottom panel shows the global chemical stratification (in hydrogen, helium, carbon, and oxygen mass fractions) of the computed representative model as a function of normalised radius. The vertical gray-dotted line, which is duplicated in other panels, roughly indicates the position of the transition between the H-rich envelope and the He-rich mantle of the star. We chose the top of this transition, where hydrogen begins to decrease inward, to mark this envelop/mantle boundary.  
The top panel shows the kernels as functions of the normalised radius of the representative g-mode ($\ell=1$, $\rm{n=-21}$) and p-mode ($\ell=1$, $\rm{n=+3}$). These modes were chosen for their period proximity with the bulk of observed g-modes and the main p-mode detected in TIC\,441725813, of which they are therefore representative. The striking feature in this panel is that both kernels sample very distinct regions of the star. The g-mode kernel is strongly confined below the envelope/mantle transition and above the convective C/O/He mixed core, which limit is the deepest chemical transition shown in the bottom panel. We point out that g-modes still propagate inside the partially mixed region that can form at the boundary of the inner convective core (contributing to about 5\% of the frequency splitting for the illustrated mode in this specific stellar model; see discussion below).
In contrast, the p-mode kernel has larger values in the upper region of the H-rich envelope, while being nearly zero in the He mantle. This suggests that, to a good approximation, the rotational splittings measured for g-modes directly sample the rotation of the star in the He-mantle, while those measured for p-modes provide the rotation rate in the H-rich envelope. To quantify further this statement, the middle panel shows the running integral of Eq.~\ref{eqn_splitting} (i.e, replacing $R$ by $r$ for the upper limit of the integral, considering further a constant rotation law $\Omega_{\rm rot}(r)\equiv$ cst) normalised by the full value of $\Delta\nu$ obtained from Eq.~(\ref{eqn_splitting}). We find that about 79\% of the splitting of g-modes comes from below the envelope, while only about 2.8\% does for the p-modes. 
To exploit this near complete separation between the internal regions respectively probed by p- and g-modes, we employ a more concise first-order formula by assuming solid-body rotation in both the mantle and the envelope (but not necessarily of the same rate). 
With solid-body rotation, the angular frequency, $\Omega_{\mathrm{rot}}$, no longer depends on the radial coordinate $r$. The frequency-splitting interval can then be expressed as
\begin{equation}
\centering
\label{rigid_rotation}
\Delta \nu_{n\ell}=\frac{\Omega_{\mathrm{rot}}}{2\pi}\left(1-C_{n\ell}\right)
\end{equation}
where $C_{n\ell}$ is the first-order solid-body rotation (or Ledoux) coefficient given by
\begin{equation}
\centering
\label{Ckl}
C_{n\ell}= \frac{\int_0^R\left[\xi^2_{h}+2\xi_{r}\xi_{h}\right]\rho r^2 \mathrm{d}r}{\int_0^R\left[\xi^2_{r}+\ell(\ell+1)\xi^2_{h}\right]\rho r^2 \mathrm{d}r}
\end{equation}
For g-modes, $C_{nl} \approx 1/[\ell(\ell+1)]$ especially in the asymptotic regime. For p-modes, $C_{n\ell}$ is usually very small compared to one and can be neglected.

\begin{figure}
   \centering
   \includegraphics[width=9cm]{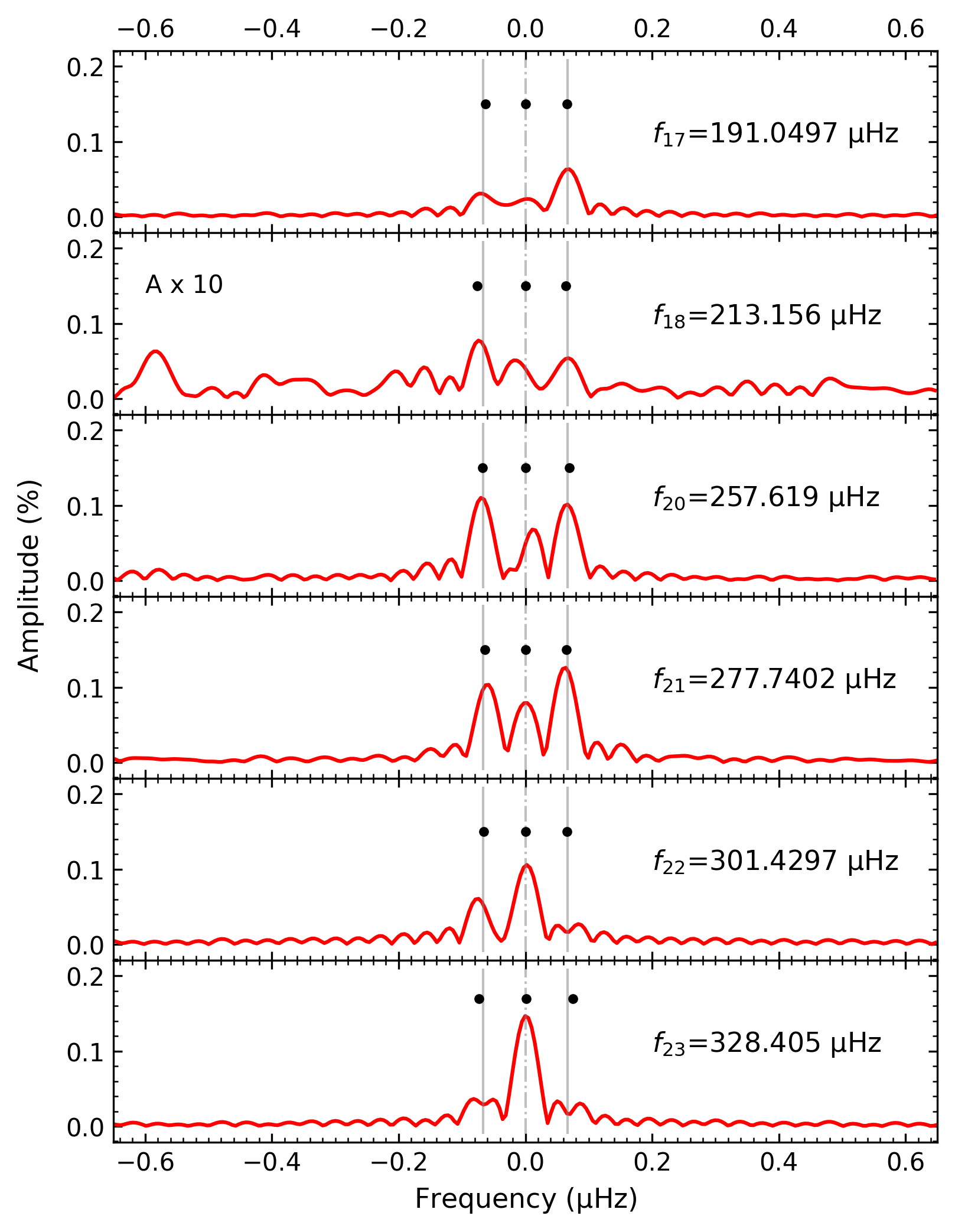}
      \caption{The 6 sets of g-mode triplets found in the first data set. The black dots indicate the effective frequencies extracted.
              }
         \label{g_triplets}
   \end{figure}

\begin{figure}
   \centering
   \includegraphics[width=9cm]{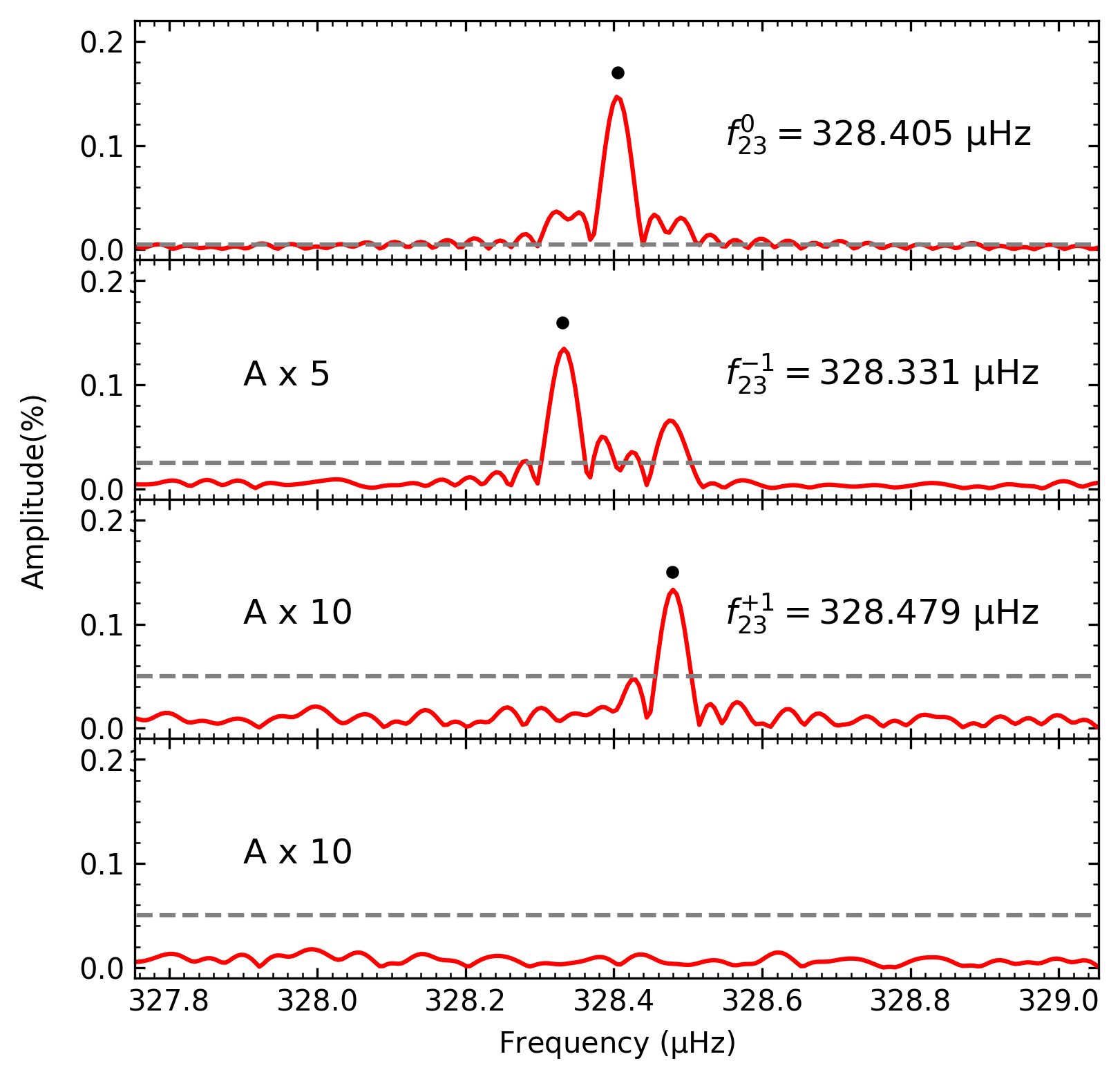}
      \caption{Three top panels: close-up views of the prewhitening process for the triplet $f_{23}$ (see Appendix~\ref{fre_mi} and Fig.~\ref{g_triplets}). The horizontal dotted line in each panel indicates 5.2 times the median noise level used as an initial significance criterion. Bottom panel: residual after prewhitening the three components of the triplet.
              }
         \label{prewhitening}
   \end{figure}

Going back to the observation of TIC\,441725813, in the g-mode region we identified 6 $\ell=1$ triplets of the same frequency splitting (Fig.~\ref{g_triplets}) present in the first data set having the longest observation time baseline and therefore the best frequency resolution. Fig.~\ref{prewhitening} provides the complete prewhitening sequence for one of these triplets, showing that there is no ambiguity in identifying the components. Those are typical signatures of the stellar rotation and we interpret them as such. The average frequency splitting derived from these triplets is $0.067\pm 0.001$ $\mu$Hz.

\begin{figure}
   \centering
   \includegraphics[width=9cm]{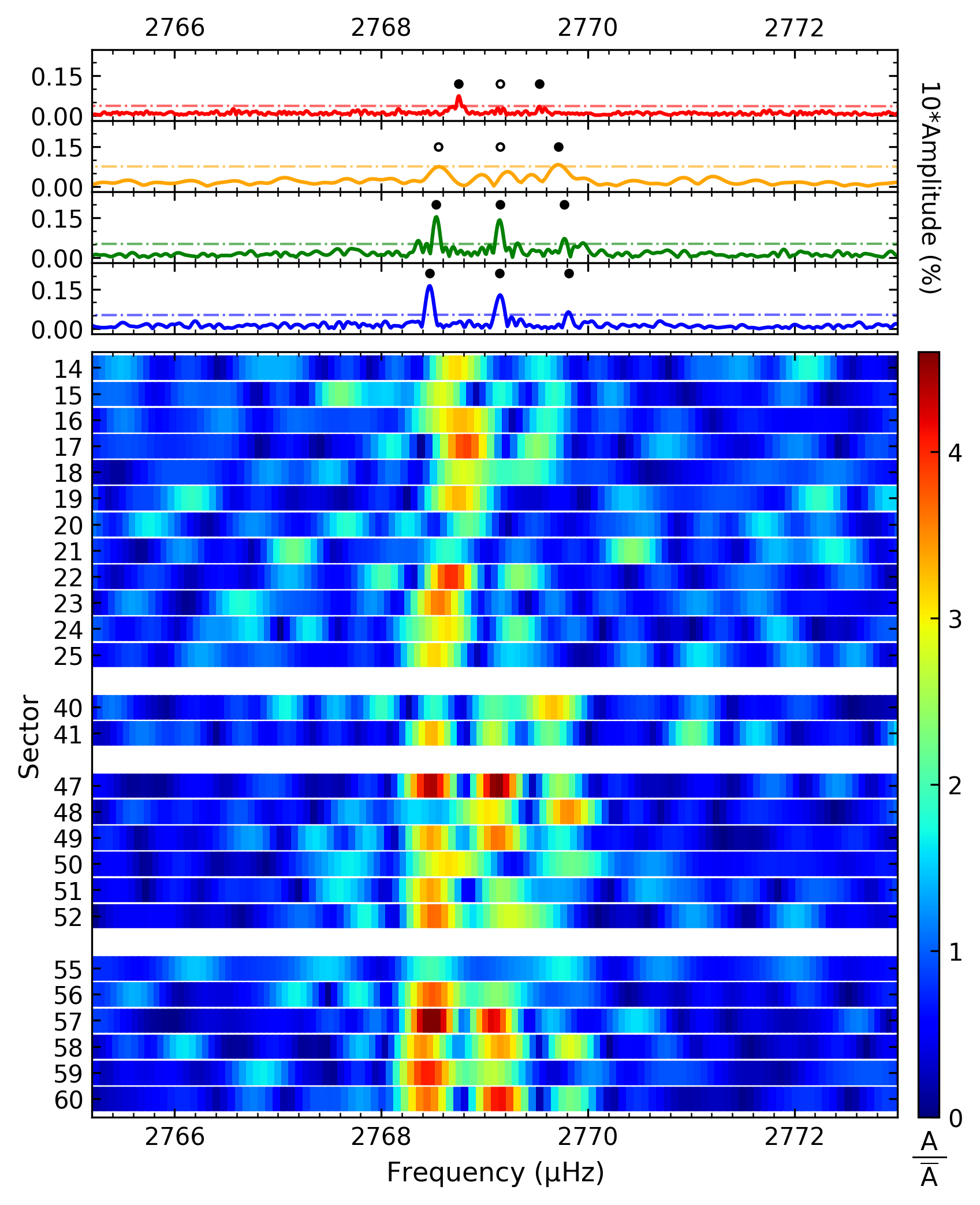}
      \caption{Close-up view of the Lomb-Scargle Periodograms for the p-mode triplet. The upper panels show the LSP of the four data sets in that region. The red horizontal line indicates our adopted detection threshold of 5.2 times the local median noise level. The black dots mark the position of signals with S/N larger than 5.2. Black open circles are possible signals just below the threshold. The lower panels show time-frequency maps built from the LSP of each sector in the same frequency range. Amplitudes are colour-encoded according to the scale shown in the colour bar on the right-hand side.
              }
         \label{p_triplets}
   \end{figure}

The second data set covers only 2 sectors and no rotation splitting could be distinguished, due to the degraded frequency resolution. However, in the third data set, we found a new triplet and confirmed 5 triplets previously seen in the first data set (see  Appendix~\ref{fre_mi}).
In addition, we found a clear frequency splitting in the p-mode region as well, in both the first, third, and fourth data sets. The top panel of Fig.~\ref{p_triplets} shows a close-up view of the LSP in that frequency range for each data set. It can be seen that there is an incomplete triplet in the first set. The central ($m=0$) frequency component is visible, but drowned in noise below our detection threshold. In the second set, only the $m=1$ frequency can be safely extracted, but in the third and fourth data sets a clear complete triplet structure emerges, leaving no doubt about its reality. We note that this p-mode multiplet exhibit clear amplitude and frequency modulation on timescales of months, which is possibly related to nonlinear mode interaction between resonant components \citep{Buchler1995}. Such phenomena have been reported by \citet{Zong2016b}, for instance in p-mode triplets of the sdB star KIC~10139564 that show amplitude and frequency modulations patterns with both short (a few hundred days) and long (over a few years) timescales. These natural nonlinear variations in amplitude and frequency induce slight uncertainties when measuring frequency spacings. The measured average frequency splitting using only the two complete p-mode triplets detected in the third and fourth data sets is $0.647\pm 0.026$\,$\mu$Hz, significantly different from the g-mode splitting. This implies from the discussion above that the outer envelope probed by the p-mode rotational splitting has a different rotation period than the deeper He-core region (made of the He mantle and the upper layers of the partially mixed He/C/O core) probed by the g-mode rotational splittings.

According to Eq.~\ref{rigid_rotation}, we estimate the average rotation period in the helium core region to be at least $85.3\pm 3.6$ days, while the outer hydrogen-rich envelope of the star rotates with an average period of $17.9\pm 0.7$ days. This means that TIC\,441725813 has radial differential rotation. It should be noted that the estimated rotation period below the envelope is possibly a lower limit due to the still significant contribution of the envelope to the rotational splitting of g-modes according to the reference kernel (see the middle panel of Fig.~\ref{kernels}). The core may actually rotate even slower, but a detailed seismic model providing individual kernels for the observed rotational triplets is required to evaluate this further. Hence, the stellar envelope of this sdB star is rotating at least $\sim 4.7$ times faster than the deeper He-core region. We recall that according to the canonical evolution scheme, the H-rich envelope is what is left from the former envelope of the red-giant progenitor, while the helium mantle, as the deeper convective helium-burning core that it surrounds, comes from the former red-giant core, itself the product of core and shell hydrogen burning on the main sequence and ascending red-giant branch. 

The fact that the envelope of this sdB star rotates significantly faster than the deeper layers is remarkable and puzzling, but not unheard of in the field as previous cases have been reported in the literature these past few years, as summarised in Table~\ref{drr}. This table lists all pulsating sdB stars, including TIC\,441725813 (this work), for which significant (more than 1.5 times difference) differential rotation is claimed, all having envelopes rotating faster. However, we believe it is important to stress that the cases for the other 5 sdB stars lack robustness and have identified weaknesses, if one pays particular attention to the frequency splittings that led to the rotation periods tagged with a question mark in Table~\ref{drr}. For instance, the p-mode multiplets in KIC\,3527751 reported by \citet{Foster2015} have been challenged by the independent analysis of \citet{zong2018} who could not detect any reliable multiplet in the p-mode region for that star. In the case of TIC~137608661, the rotation rate of the surface was deduced from rotational broadening via spectroscopy \citep{Silvotti2022}, a method that may over-estimate rotation velocity in sdB stars as found by \citet{Ma2023}. Therefore, TIC\,441725813 is arguably the most convincing case within that short list, so far. The reason why the envelope has a faster rotation is still unclear, but we note that the majority of the stars listed in Table~\ref{drr} are confirmed (or suggested, in the case of TIC\,441725813) binary systems with a close companion (short orbital period). It is therefore tantalising to attribute this faster rotation to a speed-up of the envelope caused by tidal interactions in systems that, if not yet synchronised, could be on their way to reach that state through angular momentum transport between rotation and orbital motion by internal gravity waves \citep{Goldreich1989}. This hypothesis however requires further investigations to be confirmed, both theoretical in terms of tidal synchronisation mechanisms and observational with the search for more systems like this and the confirmation of the link with binarity (the jury is still out for some objects regarding the presence or absence of a companion, including TIC\,441725813).

\begin{table*}
\centering
\caption{\label{drr} Known sdB pulsators with significantly differential radial rotation.}
\renewcommand{\arraystretch}{1.05}
\resizebox{\textwidth}{!}{
\begin{tabular}{llcllcc}
\hline
\hline
\multicolumn{1}{l}{Star} & Type          & $\mathrm{P_{orb}}$ (d) & $\mathrm{P_{core}}$ (d) & $\mathrm{P_{env}}$ (d)     & $\mathrm{P_{core}}$/$\mathrm{P_{env}}$ & Ref \\ 
\hline
\noalign{\smallskip}
TIC\,441725813             & sdB+?          & 0.28             & 85.3$\pm$3.6   & 17.9$\pm$0.7      & 4.7                  &  This work   \\
KIC\,3527751              & Single sdB?    & -             & 42.6$\pm$3.4   & 15.3$\pm$0.7?      & 2.8                  &  \cite{Foster2015}   \\
PG0048+091               & sdB+ms (wide) & ?             & 13.9$\pm$0.9?   & 4.4$\pm$0.4       & 3.2                  &  \cite{Reed2019}  \\
EPIC246023959             & sdB+M dwarf        & 0.31           & 4.6            & 2.5?                & 1.8                  &  \cite{Baran_2019}  \\ 
EPIC246683636             & sdB+M dwarf        & 0.40           & 4.2?            & 0.56              & 7.5                  &  \cite{Reed2020}  \\ 
TIC137608661             & sdB+M dwarf        & 0.30           & 4.6            & 1.3 (lower limit) & 3.5                  &  \cite{Silvotti2022}   \\ 
\hline
\end{tabular}}
\end{table*}

Finally, we point out that the rotation period derived for the deeper layers of TIC\,441725813 ($85.3\pm3.6$ days) is extremely slow, but in line with several other measurements obtained from pure g-mode sdB pulsators (see \citealt{Charpinet2018,Silvotti2022} and references therein). This value is also comparable to the typical rotation periods in the cores of red-clump stars -- i.e., core helium burning red-giant stars that are at the same evolutionary stage as sdB stars -- also measured with asteroseismology (see Fig.~9 of \citealt{mosser12} and Fig.~3 of \citealt{Mosser2024}). This similitude could be coincidental, but we prefer to interpret it as an indication that the core of TIC\,441725813 has evolved dynamically like typical red-giant stars before reaching the EHB at the onset of helium burning in the core, and has remained unaffected since then. In other words, the rotation state of the deep layers of TIC\,441725813, unlike the one we detect in the envelope, has an untouched secular origin, traced back to a red-giant progenitor following the typical dynamical evolution pattern of these stars. 


   \begin{figure}
   \centering
   \includegraphics[width=9cm]{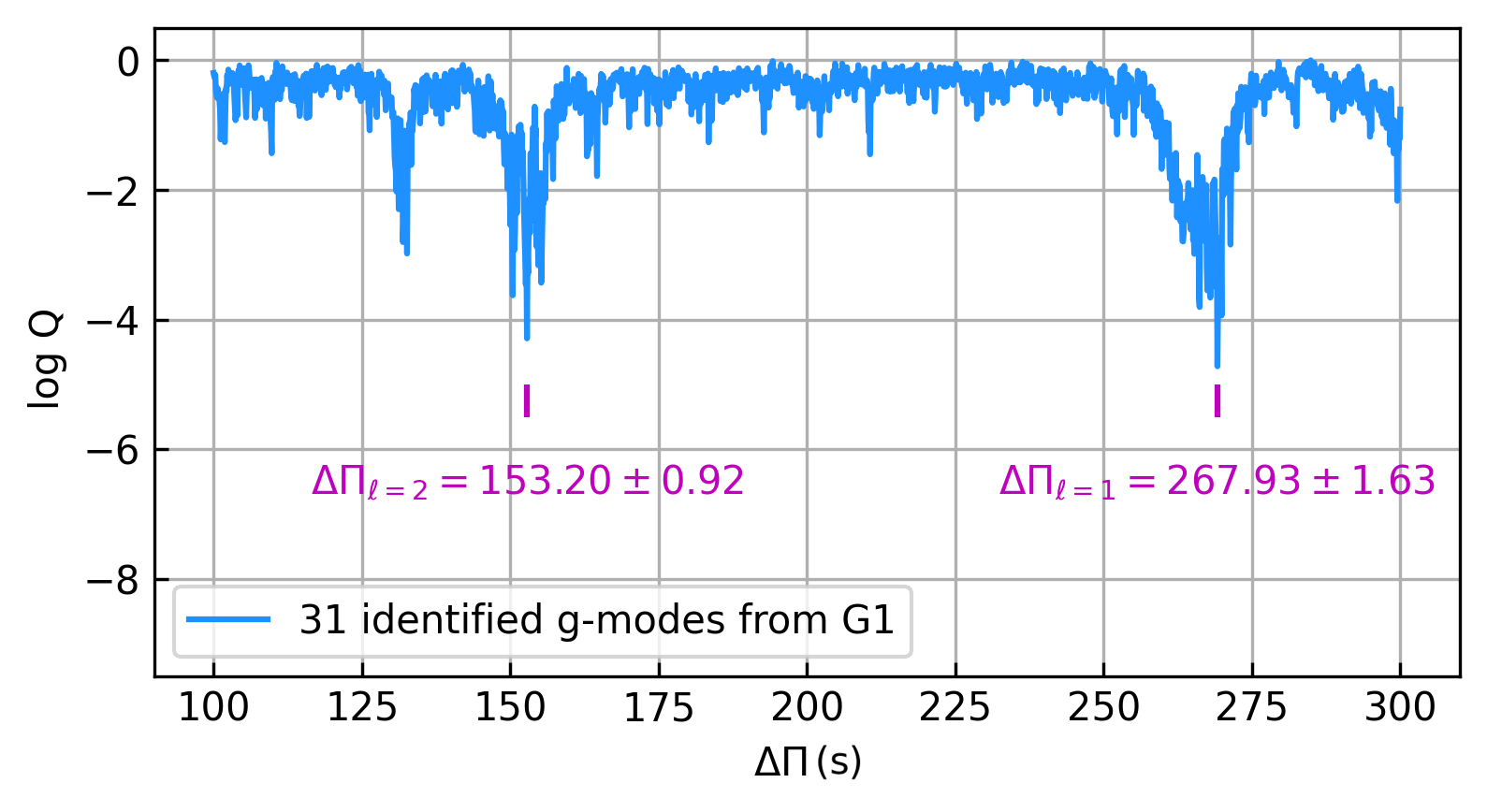}
      \caption{Results of the K-S test for 31 identified g-modes from the first data set (G1). Errors arise from the standard deviation of the results obtained from ten repeated tests.
     }
         \label{ks_test}
   \end{figure}

   \begin{figure}
   \centering
   \includegraphics[width=9cm]{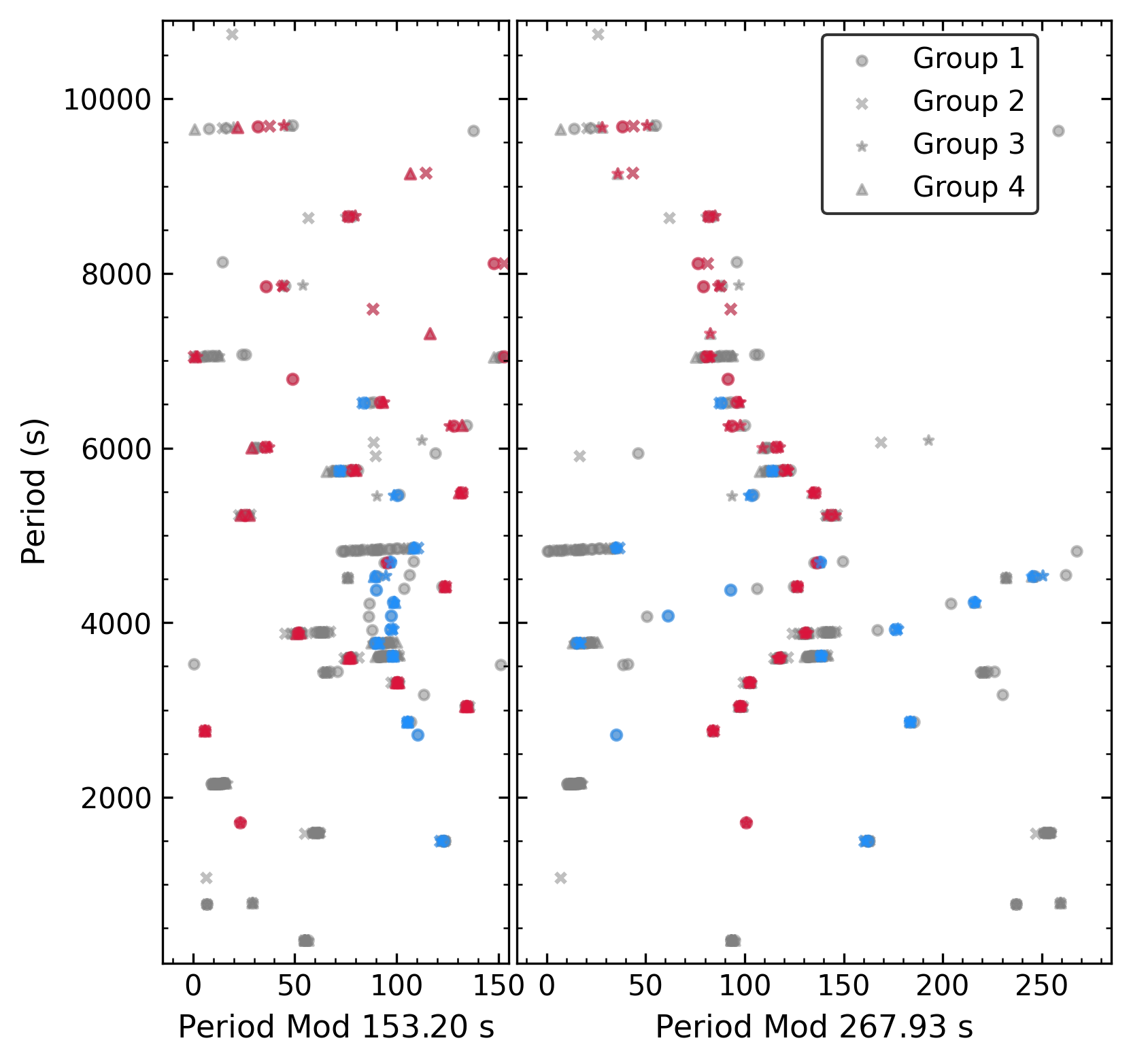}
      \caption{
      Observed frequency distribution in echelle diagrams. Various symbols represent frequencies extracted from different data sets (see the legend for details). The right-hand side panel shows the echelle diagram for a period spacing $\Delta \Pi_{\ell=1} = 267.93$\,s. The left-hand side panel is the echelle diagram for $\Delta \Pi_{\ell=2} = 153.20$\,s. Red dots are the frequencies matching the $\ell=1$ period spacing relation, while blue dots are the frequencies matching the $\ell=2$ period spacing relation. It can be seen that red and blue dots present a clear structure in the right and left echelle diagrams, respectively.}
         \label{echelle}
   \end{figure}

   \begin{figure}
   \centering
   \includegraphics[width=9cm]{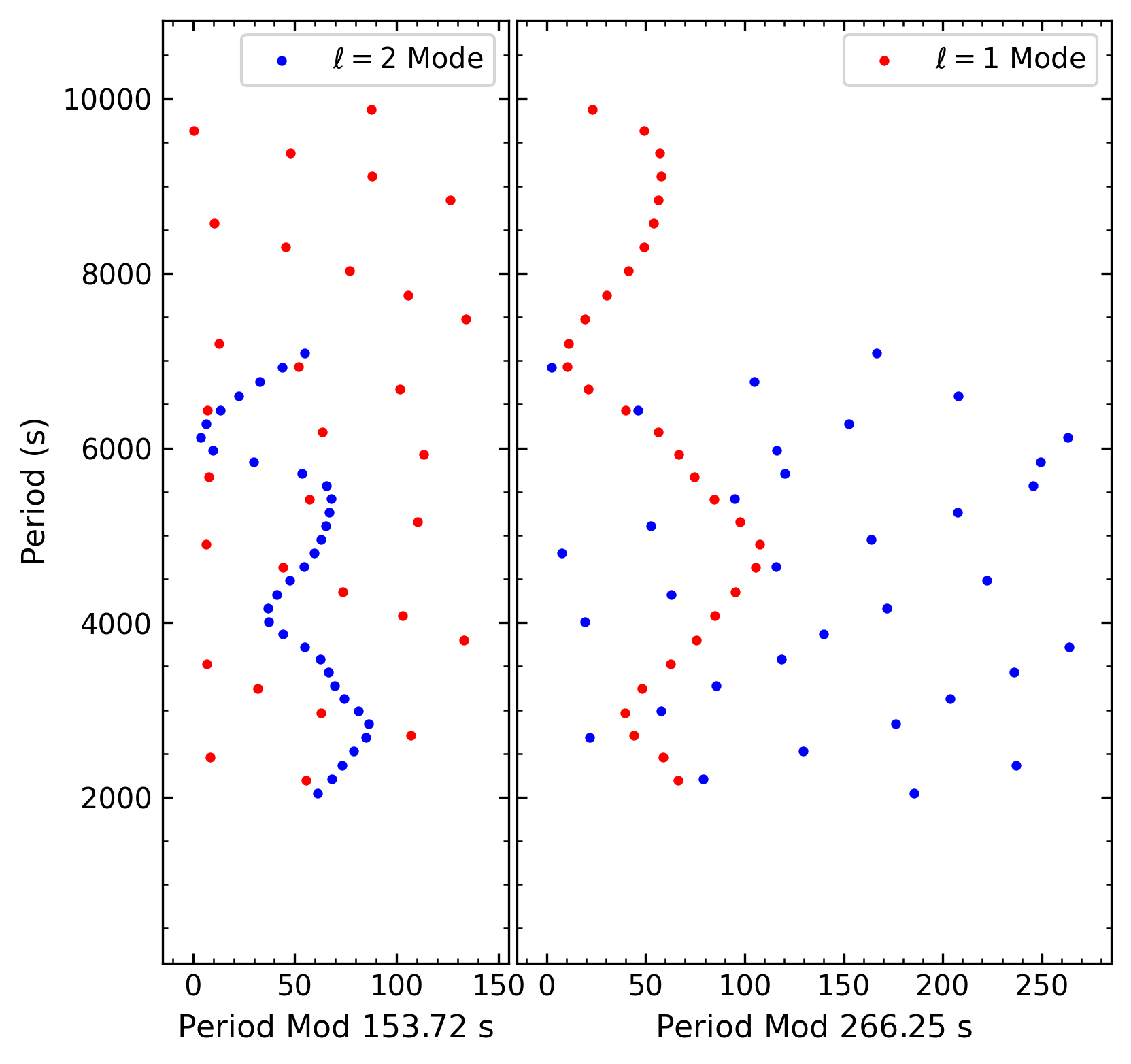}
      \caption{
      Theoretical frequency distribution in echelle diagrams for a representative sdB model. The right-hand side panel shows the echelle diagram for the computed $\ell=1$ modes (red dots) with a period spacing $\Delta \Pi_{\ell=1} = 266.25$\,s. The left-hand side panel is the echelle diagram for the computed $\ell=2$ series (blue dots) with $\Delta \Pi_{\ell=2} = 153.72$\,s. Deviation from strict asymptotic period relations is caused by structural features inside the sdB star and qualitatively shows similar patterns as observed in Fig.~\ref{echelle}. }
         \label{mode_echelle}
   \end{figure}

\subsection{Period Spacings}

We utilised the Kolmogorov–Smirnov (K-S) test to search for period spacings in the period list of the first data group. Information about the Q-value and details regarding K-S tests can be found in the work of \cite{Kawaler1988}. The results of the test are illustrated in Fig.~\ref{ks_test}. 
Two regular spacings are identified, one of $267.93\pm1.63$ s characteristic of dipolar ($\ell=1$) 
g-modes in sdB stars and another of $153.20\pm0.92$ s consistent with quadrupolar ($\ell=2$) g-modes (see below). Similar regular period spacings, ranging from 260\,s to 270\,s, have been typically reported 
in other g-mode sdB pulsators (e.g., \citealt{Foster2015,Baran_2019,2022ApJ...933..211M,Silvotti2022}).

It is well known (e.g., \citealt{Tassoul1980,Charpinet2000,Aerts2010}) that for g-mode pulsations in the asymptotic regime, modes of consecutive radial overtones tend to distribute evenly in period with a spacing of 
\begin{equation}
    \Delta \Pi_{\ell} \approx \frac{\Pi_{0}}{\sqrt{\ell(\ell+1)}}
\end{equation}
where $\Delta \Pi_{\ell} = \Pi_{\ell,n+1}-\Pi_{\ell,n}$, and $\Pi_{0}$ is the radial fundamental period.
Therefore we constructed an echelle diagram (\citealt{grec83}; see Fig.~\ref{echelle}) considering 267.93\,s as the average period spacing $\Delta\Pi$ for the $\ell=1$ series, from which we 
could identify more detected periods following the same pattern as potential $\ell=1$ modes.
With this $\ell=1$ period spacing confirmed, we searched and spotted potential $\ell=2$ modes that are expected to follow a distribution with a mean period spacing
\begin{equation}
    \Delta \Pi_{\ell=2}=\frac{\Delta \Pi_{\ell=1}}{\sqrt{3}}
\end{equation}
which, in that case, corresponds to K-S test results in Fig.~\ref{ks_test}, 153.20\,s.
Fig.~\ref{echelle} shows all detected periods from the four data sets in echelle diagrams constructed on the period spacings $\Delta \Pi = 267.93$ s and 153.20\,s, respectively.
Red points in the diagrams represent modes that are ultimately associated with $\ell=1$, while blue points represent modes tentatively identified as $\ell=2$.
In these diagrams, both $\ell=1$ and $\ell=2$ assigned periods clearly distribute along ridges since they more or less follow the asymptotic period spacing relations of their assigned $\ell$-value. These ridges however are not straight vertical lines, which indicates that the g-mode pulsation spectrum must be perturbed by features in the stellar internal structure (usually regions of rapid change of the chemical composition; see \citealt{Charpinet2000,Charpinet2013,Charpinet2014}) that make their period deviate from the pure asymptotic rule. 
A puzzling feature that appears in the left panel of Fig.~\ref{echelle} is the seemingly doubled ridge in the 3500 - 5000 s period range. Such a structure could be interpreted as the signature of rotational splitting of $\ell=2$ modes (see, e.g., \citealt{Ostensen2014}), but in that case only two components are seen and the spacing between close frequencies in those ridges is not consistent with the well identified splittings of $\ell=1$ g-modes discussed previously, at least if only consecutive $\Delta \rm m=\pm1$ modes are considered. The cavities where $\ell=1$ and 2 g-modes propagate are not very different, consequently both types of modes should reflect the same average rotation rates and have consistent splittings. The rotation hypothesis could still hold if we interpret this double-ridge structure as originating from the frequency splitting of $\ell=2$ quintuplets, with their $\rm m=0$ and $\pm 1$ components missing. In that case, we would have $\Delta \rm m=4$ between two visible components and the frequency separation would then approximately correspond to the value expected from $\ell=1$ g-mode splittings. There might be, however, another possibility : in addition to the $\ell=2$ modes, it is possible that this region of the spectrum also include several modes of higher degree ($\ell>2$). The 3500 - 5000 s period range is where the highest amplitude modes are found, i.e. where power 
is most efficiently transferred to pulsations by the driving engine. Therefore, a few strongly driven $\ell=3$ or 4 modes could possibly emerge above the detection threshold in this particular region, providing a denser set of periods, thus increasing chances of coincidental alignments in the echelle diagram. 
Nearly all frequencies belonging to this double ridge are indeed very low amplitude modes, only detected in the first (most sensitive) data set. Some of them might be the "tip of the iceberg" of the $\ell>2$ series. Unfortunately, we lack further observational evidence to confirm either of these interpretations.
Important information on the star's internal structure is contained in such period distributions and we leave the detailed seismic modelling of TIC\,441725813 based on this spectrum to the second paper of this series. 
Here, we only present, for qualitative comparison purposes, a echelle diagram of theoretical g-modes calculated for our representative model using the STELUM and PULSE codes (Fig.~\ref{mode_echelle}). It is important to emphasise that this result, similar to Fig.~\ref{kernels}, is not a best fit solution obtained after matching with the observational data of TIC\,441725813. It is sufficient, however, to emphasise the strong similarities in shape between the observed and modelled ridges. 
An important finding at this stage is the independent identification of the $\ell$-index for a significant number of pulsations modes (provided in Appendix~\ref{fre_mi}). These can be used as useful constraints in the search for a consistent seismic model of that star.

Overall, we securely identified 20 $\ell=1$ g-mode frequencies (6 sets of triplets), 15 $\ell=2$ g-mode frequencies, and 4 p-mode frequencies (containing an incomplete triplet) in the first set of data. In the second set of data, 15 $\ell=1$, 8 $\ell=2$ g-mode frequencies, and 3 p-mode frequencies were found. In the third data set, 19 g-mode frequencies with $\ell=1$ (6 sets of possible triplets), 11 g-mode frequencies with $\ell=2$, and 4 p-mode frequencies (containing a complete triplet) were found.
In the fourth data set, 17 g-mode frequencies with $\ell=1$, 6 g-mode frequencies with $\ell=2$, and 7 p-mode frequencies (also containing a complete triplet) were found.
Most of the independently identified frequencies from these four data sets are the same.
By merging repeating frequencies and grouping components associated with rotational splitting, we end up with a total of 25 $\ell=1$ and 15 $\ell=2$ independent\footnote{In this context, "independent" now refers to modes that do not belong to same rotational multiplets. In a seismic model fitting context with slow rotation involved, only the $m=0$ mode frequency carries independent information about the internal structure of the star. Other frequencies in the multiplet are determined from the m=0 frequency shifted by a fraction of the rotation rate. They are therefore not independent quantities.} (i.e., of different n and/or $\ell$-values) g-mode frequencies, and 6 independent p-mode frequencies.
The complete identification is also provided in Appendix~\ref{fre_mi}.
These frequencies will constitute the basis, in our planned follow-up paper, for the search of a best-fit asteroseismic model of TIC\,441725813 to determine its structural parameters.

\subsection{Inclination angle}
The detection of a fair number of rotation-induced triplets in the TIC\,441725813 pulsation spectrum allows us to tentatively constrain its inclination angle. Two independent methods can be used to estimate the inclination of the stellar rotation axis relative to the observer's line of sight.

\begin{figure}
   \centering
   \includegraphics[width=9cm]{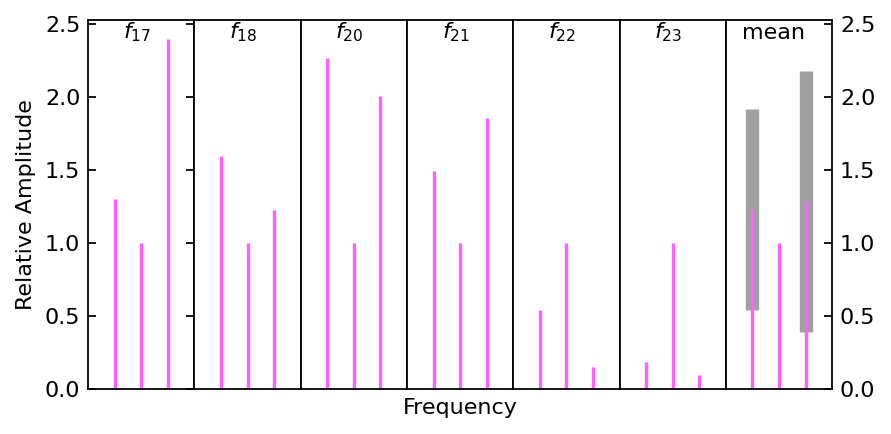}
      \caption{Relative amplitudes within six $\ell=1$ triplets detected in the first data set. The amplitude of the central component is normalized to one before computing average amplitudes, which are shown in the right-most panel. The error bars (in grey) are the standard deviations around the mean amplitudes.}
         \label{angle1}
   \end{figure}

The first approach is based on amplitude ratios between components of rotation multiplets, as illustrated in the supplementary material provided by \cite{Charpinet2011a}. In this publication, theoretical amplitude ratios were computed for different $\ell$-values and inclination angles following \citet{randall2005} and using a representative sdB model. These calculations assume that all components have the same intrinsic amplitude (hypothesis of energy equi-partition) such that the apparent amplitude variations arise solely from differing mode visibilities due to the flux modulations being integrated over the visible stellar disk for spatially unresolved stellar objects. From Figure A.5 of \cite{Charpinet2011a}, there is a clear sensitivity to the inclination of the apparent amplitude pattern within multiplets, which in principle would allow us to unambiguously determine this quantity. In practice, equi-partition of energy between components is not an obvious assumption to make for pulsators such as sdB stars that are driven by a classical $\kappa$-mechanism (as opposed to relatively short-lived, stochastically excited modes observed in Sun-like stars and red-giants). Mode amplitudes in sdB pulsators are known to vary significantly over time, and components within multiplets can be affected by long-term modulations due to weak resonant couplings \citep{Zong2016b}. Besides these complications, an argument is made in Section A.4 of \cite{Charpinet2011a} that, to first order and under slow rotation, all m-components of a given multiplet have similar eigenfunctions and, therefore, similar inertia and comparable responses to excitation through the driving mechanism. Consequently, one would expect all components to still develop roughly the same intrinsic amplitude levels and that, on average over several multiplets, despite time-varying disturbances from nonlinear mechanisms, the relative amplitude distribution would still mostly reflect the geometrical cancellation factor, hence the inclination angle. We follow that path for TIC\,441725813 with the 6 well-formed triplets identified in the data and represented in Fig.~\ref{angle1} in terms of their amplitudes relative to the normalised central component amplitude. Four out of the six triplets show similar patterns with the central component being the weakest. The two other triplets are different, with the central component being the strongest. Unfortunately, triplet detection is not reliable enough over the four data sets to discern whether some triplets may be stable or amplitude modulated, depending on the nonlinear resonant regime they are in \citep{Zong2016b,zong2018}. In this context we compute the "mean triplet" (right-most panel in Fig.~\ref{angle1}) from the 6 available triplets without distinction, leading to a rather large standard deviation that translates into a rather soft constraint on the inclination angle. Still, Comparing these numbers and considering the error in the average amplitude ratio, we estimate the rotation inclination of TIC\,441725813 to be ${60^\circ}^{+10^\circ}_{-20^\circ}$.

\begin{figure}
   \centering
   \includegraphics[width=8cm]{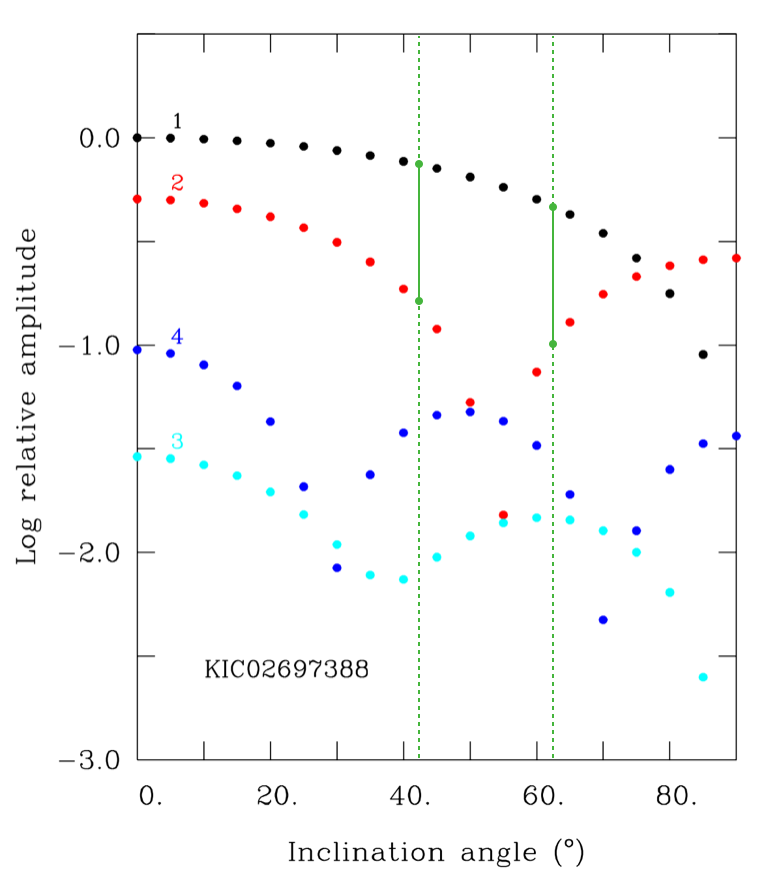}
      \caption{Relative amplitude (in logarithmic scale) for g-mode oscillations as a function of inclination angle and degree $\ell$ (assuming $m=0$). Degrees $\ell=1$ (black), 2 (red), 3 (cyan), and 4 (blue) are represented. These visibility functions were evaluated by \cite{Charpinet2011b} for a sdB model with parameters representative of KIC\,2697388 observed in the {\sl Kepler} band-pass. For simplicity, we adopt the same model for TIC\,441725813. The Green lines indicate the constraints we obtain for that star.}
         \label{angle2}
   \end{figure}

The second approach to estimate the inclination angle is by calculating observed average amplitude ratios between modes of different degrees $\ell$ and comparing them to visibility calculations, as illustrated in \cite{Charpinet2011b}. Compared to the previous method, this approach has an additional underlying assumption that the modes considered are $m=0$, 
which we cannot guarantee in most cases. Therefore the following results are at best indicative.  
Fig.~\ref{angle2} shows the theoretical relative amplitudes calculated by \cite{Charpinet2011b} for a model representative of the sdB star KIC\,2697388 observed by the Kepler instrument. These calculations were tuned for the atmospheric parameters of that star (which are not very different from those of TIC\,441725813) and the Kepler band-pass, which differs from the redder TESS band-pass. However, since both Kepler and TESS photometry sample the tail of the spectral energy distribution for hot sdB stars, we deemed it unnecessary to recompute a tuned model for TIC\,441725813 that would not change significantly the results, in particular considering all uncertain assumptions that anyway come into play. 
We calculated the average amplitude ratio between the identified $\ell=1$ and $\ell=2$ modes of TIC\,441725813, leading to $\log(\bar A_{\ell=1}/\bar A_{\ell=2}) \approx 0.687$. The green lines in Fig.~\ref{angle2} show at which inclination angles this ratio occurs. Two solutions are suggested, implying an inclination around $\sim$ $40^{\circ}$ or $\sim$ $60^{\circ}$. The second solution appears to be better in line with the inclination estimated from the previous method, although uncertainties are high, once again.

To sum up, combining the results obtained independently from the two methods, the suggested inclination for TIC\,441725813 is about $\sim$ $60^{\circ}$. However, limited by uncertainties, we cannot provide a more stringent constraint than this with the available data.

\section{Summary and conclusion}
We analysed the photometric data obtained by TESS (more than 670 days of cumulated monitoring) for the pulsating sdB star TIC\,441725813. From these, we established that it is a hybrid pulsator showing both g- and p-modes. The spectroscopy available to us for that star indeed places it, with $\mathrm{T_{\rm eff}} = 27827\pm 177\,\mathrm{K}$ and $\log g = 5.463\pm 0.028$, well within the g-mode instability region where V1093 Hya stars are typically found, and at the low-temperature edge of the region usually harboring hybrid sdB pulsators.

The detailed study of the light curves in Fourier space revealed a particularly rich spectrum containing, from low to high frequencies, the possible orbital signature of a companion, a furnished g-mode pulsation spectrum with signatures of rotational splitting, and a few p-modes showing also the signature of rotation. The latter allowed us to estimate the rotation periods of both the core and the outer envelope, from the measured frequency splittings of the two types of modes. We find that the core rotates very slowly with a period of at least $85.3\pm 3.6$ days (in line with rotation rates measured from asteroseismology in other sdB pulsators and in the cores of Red Clump stars), while the rotation period of the outer envelope is $17.9\pm 0.7$ days. We therefore find that the envelope rotates at least about 4.7 times faster than the inner core of the star. This raises the question of whether binarity, through tidal interactions, has a role to play in explaining this differential rotation. Important in this context, we found strong indications, from both a weak photometric signal extracted from the TESS data and from a significant spectroscopic radial velocity jitter in our available spectra, that TIC\,441725813 may possibly be a short period sdB+WD binary system, with an orbital period of about 6.7 hours. Additional spectroscopic data are however needed to confirm this interpretation. 

Exploring further the pulsation spectrum of TIC\,441725813, we identified the degree $\ell$ of several pulsation modes by combining rotational splitting information (when available) and asymptotic period spacings for adjacent g-modes. We established that the $\ell=1$ and $\ell=2$ modes have a mean period spacing, $\Delta \Pi$, of $267.93\pm1.63$\,s and $153.20\pm0.92$\,s, respectively.
Without counting splitted frequency components, we ended up with the identification of 25 $\ell=1$ g-mode frequencies, 15 $\ell=2$ g-mode frequencies, and 6 p-mode frequencies.
From those, using two independent methods, we also attempted to estimate the inclination angle of the rotation/pulsation axis of the star relative to the observer, suggesting an inclination of $\sim$ $60^{\circ}$ but with large uncertainties. 

All the above information extracted from the photometric and spectroscopic observation of TIC\,441725813 paves the way for a detailed asteroseismic analysis of this star to
determine its structural parameters. This important modelling effort will be reported in a follow-up paper.

\begin{acknowledgements}
WS and SC acknowledge financial support from the Centre National d’\'Etudes Spatiales (CNES, France) and partially from the Agence Nationale de la Recherche (ANR, France) under grant ANR-17-CE31-0018.

M.L. acknowledges funding from the Deutsche Forschungsgemeinschaft (grant LA 4383/4-1).

W.Z. acknowledges the support from the National Natural Science Foundation of China through grants 12273002, 12090040/42 and the science research grants from the China Manned Space Project.
\end{acknowledgements}

%
   \bibliographystyle{aa} 
   \bibliography{aanda} 
%

\begin{appendix} 
{\onecolumn
\section{List of extracted frequencies interpreted as independent signals for TIC\,441725813}
\begin{table*}[h]
    \centering
    \label{fre_mi}
    \renewcommand{\arraystretch}{0.6}
    \resizebox{\textwidth}{!}{
    \begin{tabular}{l|llll|llll|cccc}
    \hline
    Id. & \multicolumn{4}{c|}{Frequency ($\mu Hz$)} & \multicolumn{4}{c|}{Period (s)} & \multicolumn{4}{c}{SNR \footnotemark[1]} \\
    \hline
    & \multicolumn{1}{c}{G1} & \multicolumn{1}{c}{G2} & \multicolumn{1}{c}{G3} & \multicolumn{1}{c|}{G4} & \multicolumn{1}{c}{G1} & \multicolumn{1}{c}{G2} &\multicolumn{1}{c}{G3} & \multicolumn{1}{c|}{G4} & \multicolumn{1}{c}{G1} &\multicolumn{1}{c}{G2} & \multicolumn{1}{c}{G3}&\multicolumn{1}{c}{G4}\\
    \hline
    \multicolumn{13}{c}{Orbital signals} \\
    \hline
    $f_{a}$&40.928(3)&…&…&…&24432(2)&…&…&…&5.9&…&…&…\\
    $f_{b}$&41.603(3)&41.59(1)&…&41.584(5)&24036(2)&24039(8)&…&24047(2)&5.6&7.8&…&10.1\\
    \hline
    \multicolumn{13}{c}{$g$-modes with $\ell=1$} \\
    \hline
    $f_{1}$&…&…&72.770(6)&72.774(6)&…&…&13741(1)&13741(1)&…&…&6.1&7.4\\
    $f_{2}$&…&93.08(2)&…&…&…&10743(2)&…&…&…&6.04&…&…\\
    $f_{3}$&103.267(2)&103.20(2)&103.132(4)&103.374(4)&9683.5(2)&9689(2)&9696.2(4)&9673.5(4)&9.6&5.5&8.5&12.8\\
    $f_{4}$&…&109.25(2)&…&109.345(6)&…&9153(1)&…&9145.3(5)&…&5.9&…&7.7\\
    $f_{5}$&115.533(1)&115.533(9)&115.489(3)&115.529(5)&8655.4(1)&8655.5(7)&8658.7(2)&8655.8(5)&15.2&12.3&13.4&7.1\\
    $f_{6}$&123.242(3)&123.16(2)&…&…&8114.0(2)&8119(1)&…&…&5.4&5.5&…&…\\
    $f_{7}$&127.404(2)&127.266(8)&127.279(4)&…&7849.0(1)&7857.5(5)&7856.7(2)&…&9.0&13.6&10.3&…\\
    $f_{8}$&…&131.66(1)&…&…&…&7594(1)&…&…&…&6.7&…&…\\
    $f_{9}$&…&…&…&136.675(6)&…&…&…&7316.6(3)&…&…&…&8.0\\
    &&&&&&&&&&&&\\
    $f_{10}^{-1}$&…&…&141.797(2)&141.781(3)&…&…&7052.3(1)&7053.1(1)&…&…&15.9&15.8\\
    $f_{10}$&141.9095(5)&141.88(1)&141.8617(4)&141.8781(8)&7046.74(2)&7047.7(6)&7049.12(1)&7048.30(3)&42.8&9.3&112.1&61.2\\
    $f_{10}^{+1}$&…&…&141.939(1)&141.937(4)&…&…&7045.26(8)&7045.3(2)&…&…&22.8&14.5\\
    &&&&&&&&&&&&\\
    $f_{11}$&147.284(3)&…&…&…&6789.5(1)&…&…&…&6.5&…&…&…\\
    $f_{12}$&153.224(1)&…&153.186(6)&153.202(6)&6526.36(5)&…&6527.9(2)&6527.2(2)&16.3&…&6.1&8.0\\
    $f_{13}$&159.847(1)&…&159.895(7)&159.744(6)&6255.97(3)&…&6254.0(2)&6260.0(2)&19.7&…&5.7&7.2\\
    $f_{14}$&166.383(2)&166.37(1)&166.335(5)&166.569(5)&6010.21(6)&6010.6(5)&6011.9(2)&6003.4(1)&10.6&7.4&7.6&10.2\\
    $f_{15}$&174.0268(1)&173.988(1)&173.9688(2)&173.9645(2)&5746.240(4)&5747.52(3)&5748.156(9)&5748.29(1)&145.7&104.3&145.5&154.4\\
    $f_{16}$&182.014(1)&182.01(1)&182.035(4)&182.061(7)&5494.06(3)&5494.0(3)&5493.4(1)&5492.6(2)&17.6&9.8&10.5&7.3\\
    &&&&&&&&&&&&\\
    $f_{17}^{-1}$&190.9852(5)&…&190.977(1)&190.973(1)&5236.00(1)&…&5236.20(3)&5236.33(3)&39.9&…&35.0&37.7\\
    $f_{17}^{0}$&191.0479(5)&…&…&…&5234.28(1)&…&…&…&30.75&…&…&…\\
    $f_{17}^{+1}$&191.1138(2)&…&191.1227(8)&191.125(1)&5232.485(7)&…&5232.24(2)&5232.15(2)&73.5&…&46.5&50.8\\
    &&&&&&&&&&&&\\
    $f_{18}^{-1}$&213.079(2)&…&213.091(6)&…&4693.07(4)&…&4692.8(1)&…&9.9&…&6.3&…\\
    $f_{18}^{0}$&213.156(3)&…&213.160(5)&…&4691.39(6)&…&4691.3(1)&…&6.2&…&7.5&…\\
    $f_{18}^{+1}$&213.220(3)&…&…&…&4689.98(5)&…&…&…&7.6&…&…&…\\
    &&&&&&&&&&&&\\
    $f_{19}$&226.5885(5)&226.578(8)&226.591(2)&226.586(2)&4413.28(1)&4413.4(1)&4413.23(3)&4413.32(4)&36.7&14.8&20.6&22.9\\
    &&&&&&&&&&&&\\
    $f_{20}^{-1}$&257.5511(1)&…&257.5562(5)&257.5617(5)&3882.722(4)&…&3882.647(7)&3882.564(7)&141.1&…&86.3&156.8\\
    $f_{20}^{0}$&257.6190(2)&…&257.614(2)&…&3881.700(5)&…&3881.77(2)&…&62.3&…&25.8&…\\
    $f_{20}^{+1}$&257.6884(2)&…&257.6977(5)&257.6786(5)&3880.656(2)&…&3880.516(8)&3880.802(9)&124.9&…&70.5&79.7\\
    &&&&&&&&&&&&\\
    $f_{21}^{-1}$&277.6764(1)&…&277.6874(4)&277.6866(5)&3601.314(1)&…&3601.171(5)&3601.182(7)&134.5&…&93.3&99.0\\
    $f_{21}^{0}$&277.7402(2)&…&277.743(6)&277.749(3)&3600.486(3)&…&3600.44(7)&3600.36(3)&90.4&…&6.5&46.1\\
    $f_{21}^{+1}$&277.8048(1)&…&277.7958(2)&277.7972(5)&3599.650(1)&…&3599.765(4)&3599.748(6)&167.2&…&118.9&130.5\\
    &&&&&&&&&&&&\\
    $f_{22}^{-1}$&301.3636(2)&…&301.353(1)&301.358(1)&3318.250(2)&…&3318.36(1)&3318.30(1)&81.3&…&38.1&51.5\\
    $f_{22}^{0}$&301.4297(1)&301.412(1)&301.4359(4)&301.4243(5)&3317.522(1)&3317.71(1)&3317.454(5)&3317.583(6)&151.4&86.6&88.0&104.1\\
    $f_{22}^{+1}$&301.4956(9)&…&301.497(3)&301.482(5)&3316.797(9)&…&3316.78(3)&3316.94(5)&22.5&…&13.9&11.1\\
    &&&&&&&&&&&&\\
    $f_{23}^{-1}$&328.3310(5)&…&328.323(1)&328.327(2)&3045.706(4)&…&3045.78(1)&3045.73(1)&40.6&…&24.6&28.3\\
    $f_{23}^{0}$&328.4050(1)&328.408(1)&328.4038(2)&328.4055(4)&3045.0206(8)&3044.99(1)&3045.031(2)&3045.015(3)&219.0&67.3&128.4&149.1\\
    $f_{23}^{+1}$&328.479(1)&…&328.475(2)&328.477(2)&3044.332(8)&…&3044.36(1)&3044.34(2)&20.6&…&19.3&26.6\\
    &&&&&&&&&&&&\\
    $f_{24}$&361.890(1)&361.87(1)&361.885(3)&361.894(5)&2763.267(8)&2763.3(1)&2763.30(2)&2763.23(3)&18.6&7.4&12.2&12.6\\
    $f_{25}$&585.321(2)&…&585.324(7)&…&1708.463(6)&…&1708.45(2)&…&8.8&…&5.6&…\\
    \hline
    \multicolumn{13}{c}{$g$-modes with $\ell=2$} \\
    \hline
    $f_{1}^{*}$&153.413(3)&153.42(1)&…&…&6518.3(1)&6517.7(6)&…&…&6.6&7.4&…&…\\
    $f_{2}^{*}$&174.209(1)&174.207(5)&174.183(2)&…&5740.22(5)&5741.0(1)&5740.90(5)&…&12.1&9.1&21.3&…\\
    $f_{3}^{*}$&183.076(2)&…&183.126(7)&…&5462.18(6)&…&5460.7(2)&…&9.6&…&5.5&…\\
    $f_{4}^{*}$&205.856(2)&205.78(1)&205.852(2)&205.869(4)&4857.74(5)&4859.3(3)&4857.84(5)&4857.43(9)&8.2&8.1&17.1&12.8\\
    $f_{5}^{*}$&213.079(2)&…&213.091(6)&…&4693.07(4)&…&4692.8(1)&…&9.9&…&6.3&…\\
    $f_{6}^{*}$&220.629(3)&…&220.397(6)&220.661(6)&4532.48(6)&…&4537.2(1)&4531.8(1)&6.7&…&6.2&9.6\\
    $f_{7}^{*}$&228.333(4)&…&…&…&4379.55(6)&…&…&…&5.7&…&…&…\\
    $f_{8}^{*}$&236.148(3)&…&236.127(4)&236.115(6)&4234.63(6)&…&4235.00(7)&4235.21(1)&5.7&…&8.8&8.5\\
    $f_{9}^{*}$&245.078(2)&…&…&…&4080.32(2)&…&…&…&10.9&…&…&…\\
    $f_{10}^{*}$&254.648(3)&254.60(2)&254.556(5)&…&3926.98(3)&3927.6(3)&3928.39(8)&…&7.7&6.2&7.5&…\\
    $f_{11}^{*}$&265.5118(4)&265.398(3)&265.397(1)&265.418(1)&3766.311(6)&3767.91(4)&3767.92(1)&3767.63(1)&48.6&35.3&39.7&55.5\\
    $f_{12}^{*}$&276.125(3)&276.122(3)&276.1189(8)&276.0976(9)&3621.54(4)&3621.58(3)&3621.62(1)&3621.90(1)&6.3&37.5&49.0&63.5\\
    $f_{13}^{*}$&349.299(1)&349.31(1)&349.301(4)&349.307(5)&2862.87(1)&2862.7(1)&2862.85(3)&2862.81(4)&15.5&7.3&10.0&12.0\\
    $f_{14}^{*}$&368.375(3)&…&…&…&2714.61(2)&…&…&…&5.9&…&…&…\\
    $f_{15}^{*}$&665.915(2)&666.55(1)&666.499(6)&665.931(5)&1501.691(5)&1500.25(2)&1500.37(1)&1501.66(3)&7.9&9.0&6.2&6.5\\
    \hline
    \multicolumn{13}{c}{$p$-modes} \\
    \hline
    $f_{\uppercase\expandafter{\romannumeral1}}$&1257.804(1)&1257.77(1)&1257.780(5)&1257.783(3)&795.0358(9)&795.05(1)&795.051(3)&795.049(1)&12.8&6.8&8.1&20.3\\
    $f_{\uppercase\expandafter{\romannumeral2}}$&1293.780(2)&1293.77(1)&1293.789(7)&1293.943(8)&772.928(1)&772.932(6)&772.923(4)&772.832(7)&9.0&10.3&5.8&5.9\\
    $f_{\uppercase\expandafter{\romannumeral3}}$&2755.541(3)&…&2755.546(5)&2755.533(5)&362.9051(3)&…&362.9043(6)&362.9060(7)&6.6&…&8.3&10.8\\
    &&&&&&&&&&&&\\
    $f_{\uppercase\expandafter{\romannumeral4}}^{-1}$&2768.750(2)&…&2768.530(3)&2768.471(3)&361.1737(2)&…&361.2024(3)&361.2101(4)&9.8&…&14.5&17.5\\
    $f_{\uppercase\expandafter{\romannumeral4}}^{0}$&…&…&2769.148(3)&2769.145(2)&…&…&361.1218(3)&361.1219(6)&…&…&13.8&13.0\\
    $f_{\uppercase\expandafter{\romannumeral4}}^{+1}$&2769.529(4)&2769.71(2)&2769.771(6)&2769.815(5)&361.0721(4)&361.048(2)&361.0406(7)&361.0348(7)&5.2&5.6&6.7&6.7\\
    &&&&&&&&&&&&\\
    $f_{\uppercase\expandafter{\romannumeral5}}$&...&...&...&2797.017(6)&...&...&...&357.523(8)&...&...&...&5.7\\
    $f_{\uppercase\expandafter{\romannumeral6}}$&...&...&...&3661.141(6)&...&...&...&273.138(5)&...&...&...&5.6\\
    $f_{\uppercase\expandafter{\romannumeral7}}$&...&...&...&3678.591(6)&...&...&...&271.843(4)&...&...&...&6.1\\
    \hline
\end{tabular}

    }
    \footnotemark[1]\parbox{\linewidth}{\raggedright\fontsize{8}{10}\selectfont For the first data set (G1), Amplitude = $0.96 * 10^{-3}$ * S/N (\%); For G2, Amplitude = $1.95 * 10^{-3}$ * S/N (\%); For G3, Amplitude = $1.48 * 10^{-3}$ * S/N (\%); For G4, Amplitude = $1.74 * 10^{-3}$ * S/N (\%).
    }
\end{table*}
}
\end{appendix}

\end{document}